\documentclass[%
 reprint,
 amsmath,amssymb,
 aps,
]{revtex4-2}

\usepackage{graphicx}
\usepackage{dcolumn}
\usepackage{bm}
\usepackage{amsmath}
\usepackage[version=4]{mhchem}
\usepackage{siunitx}
\usepackage{tabularx}
\usepackage{optidef}

\usepackage{subcaption}

\usepackage{algorithm}
\usepackage{algpseudocode}
\DeclareMathOperator*{\argmax}{\arg\!\max}

\algblock{Input}{EndInput}
\algnotext{EndInput}
\algblock{Output}{EndOutput}
\algnotext{EndOutput}

\usepackage{xpatch}
\makeatletter
\xpatchcmd{\algorithmic}{\itemsep\z@}{\itemsep=1ex plus1pt}{}{}
\makeatother

\newcommand{\mycomment}[1]{}
\newcommand{\trans}{^\mathsf{T}}

\begin{document}

\preprint{APS/123-QED}

\title{An optimization framework for analyzing\\ nonlinear stability due to sparse finite-amplitude perturbations}

\author{A. Leonid Heide and Maziar S. Hemati}
 \affiliation{Aerospace Engineering and Mechanics, University of Minnesota} 

\date{\today}

\begin{abstract}

Recent works have established the utility of sparsity-promoting norms for extracting spatially-localized instability mechanisms in fluid flows, with possible implications for flow control.
However, these prior works have focused on linear dynamics of infinitesimal perturbations about a given baseflow.
In this paper, we propose an optimization framework for computing sparse finite-amplitude perturbations that maximize transient growth in nonlinear systems.
A variational approach is used to derive the first-order necessary conditions for optimality, which form the basis of our iterative direct-adjoint looping numerical solution algorithm.
When applied to a reduced-order model of a sinusoidal shear flow at $Re=20$, our framework identifies that energy injection into a single vortical mode yields comparable energy amplification as the non-sparse optimal solution with energy distributed across all modes.
Energy injection into three additional modes results in an identical transient growth as the non-sparse case.
Subsequent analysis of the dynamic response of the flow establishes that these sparse optimal perturbations trigger many of the same nonlinear modal interactions that give rise to transient growth when all modes are perturbed in an optimal manner.
It is also observed that as perturbation amplitude is increased, the maximum transient growth is achieved at an earlier time.
Our results highlight the power of the proposed
optimization framework for revealing dominant nonlinear modal interactions and sparse perturbation
mechanisms for transient growth and instability in fluid flows.
We anticipate the approach will be a useful tool in guiding the design of flow control strategies in the future.

\end{abstract}

\maketitle

\section{\label{sec:Introduction}Introduction}
Transient growth refers to the amplification of perturbations over a finite-time horizon, and is an important mechanism for instability in fluids systems~\cite{schmidBook}.
Perturbation energy can grow over finite transient time-horizons, even when the dynamics are linear and stable.
This is the basis of non-modal stability theory~\cite{schmidARFM2007}.
The transient growth phenomenon arises due to a large-degree of non-normality in the linear operator that governs the dynamics.
Within the context of incompressible flows, 
transient growth of the linearized dynamics is often a necessary condition for instability of perturbations governed by the associated nonlinear  dynamics~\cite{schmidBook}.

Worst-case analysis is a common feature of transient growth studies.
Among all admissible finite-amplitude perturbations, we are most interested in the one that maximizes some norm (e.g.,~kinetic energy) over a given time horizon.
Such a perturbation is referred to as an \emph{optimal perturbation}.
For linear perturbation dynamics, the optimization problem for a \emph{linear optimal perturbation~(LOP)} reduces to maximizing a Rayleigh quotient, which corresponds to solving an eigenvalue problem~\cite{schmidAMR2014,schmidARFM2007,schmidBook}.
For nonlinear dynamics, a \emph{nonlinear optimal perturbation~(NLOP)} of a prescribed amplitude can be computed by solving the associated optimization problem using the calculus of variations.
This is the basis of the so-called nonlinear non-modal stability theory~\cite{kerswell2014,kerswellAnnRev2018}.
An optimization over the perturbation amplitude and time-horizon can be used to determine the \emph{minimal seed} for instability in a nonlinear system.
An overview of the details of these approaches will be given in Section~\ref{sec:NLOP}.
For additional details, the reader is pointed to several excellent review articles~\cite{schmidARFM2007,schmidAMR2014,kerswell2014,kerswellAnnRev2018}.

In general, optimal perturbations that maximize transient growth in fluids systems are spatially extended and lack sparsity---i.e.,~many coupled flow variables contribute to an optimal perturbation, rather than a smaller (sparse) subset of flow variables.
Yet, it is often of interest to identify sparse and spatially-localized optimal perturbations.
Sparse and spatially-localized optimal perturbations would reveal specific spatial locations where perturbations of a specific quantity would dominate in driving baseflow instability. Similarly, obtaining sparse optimal perturbations in a suitable modal basis would reveal dominant modes and associated physical processes.
Such solutions could also provide guidance on actuator placement and inform flow control strategies by revealing locations and quantities where the effect of control can be most pronounced~\cite{bhattacharjeeTCFD2020,skeneJFM2022,tamilselvamAIAA2022}. 

Recent works have leveraged sparsity promoting $\ell_1$-norms to sparsify the optimal forcing modes identified by resolvent analysis~\cite{Skene2022,Lopez-Doriga2023}.
An alternative approach based on $p$-norm maximization was proposed in~\cite{Foures2013}.
In particular, it was shown that maximizing the $\infty$-norm of energy amplification over a given time-horizon revealed spatially-localized perturbations.
All of these works successfully demonstrated that sparse and spatially-localized perturbations can be identified by considering the linear dynamics of perturbations about a given baseflow.
However, the assumption of linear dynamics implicitly assumes that all pertubrations are of infinitesimal magnitude.
In reality, finite-amplitude effects can play an important role in energy amplification driven by the \emph{nonlinear} fluid dynamics.

In this paper, we present a framework for computing sparse finite-amplitude optimal perturbations that maximize transient growth of perturbation energy in nonlinear systems.
The approach can also be extended to compute spatially-localized perturbations in systems governed by partial differential equations.
The proposed optimization framework augments the standard transient growth objective function in NLOP analysis with a sparsity-promoting $\ell_1$-norm on the initial perturbation.
We formulate a variational method to solve the associated sparse NLOP problem based on a direct-adjoint looping~(DAL) procedure.
The DAL algorithm can be used to determine sparse optimal perturbations associated with a given (finite) perturbation amplitude.
Gridding over pertubration amplitude can be used to identify the finite amplitude effects and nonlinear flow interactions that give rise to transient growth.
Alternatively, bisection over the perturbation amplitude can be used to identify sparse  finite-amplitude perturbations that destabilize the flow (i.e., sparse minimal seeds).

The proposed framework is demonstrated on two finite-dimensional nonlinear systems 
based on the incompressible Navier-Stokes equations:
(1)~a simple 2-state model that exhibits important features of the incompressible Navier-Stokes equations, and (2)~a reduced-order model of a sinusoidal shear flow.
The 2-state model is amenable to graphical demonstrations and is used to highlight important features of the sparse NLOP problem and associated solutions.
The reduced-order model of sinusoidal shear flow provides a non-trivial benchmark and highlights the utility of the proposed method in extracting important instability mechanisms. In particular, the proposed method identifies that perturbation of a single mode yields transient energy growth comparable to the non-sparse optimal solution. This sparse perturbation excites the most dynamically significant mode without any prior knowledge of its physical and dynamic importance. Furthermore, the framework is also used to identify sparse energy-maximizing perturbations of several modes, which provides insight into coupling between the modes. These behaviors are consistent with the mechanisms observed during the transition to turbulence which the model was constructed to predict~\cite{Moehlis2004}. 

The paper is organized as follows:
In Section~\ref{sec:NLOP}, we review the standard NLOP optimization problem and the associated direct-adjoint looping~(DAL) solution algorithm.
In Section~\ref{sec:spNLOP}, we introduce the proposed sparse NLOP optimization problem and a corresponding DAL solution algorithm.
The approach is demonstrated on two finite-dimensional systems with results presented in Sections ~\ref{sec:AnExample} and \ref{sect:model_overview}.
Section~\ref{sec:conclusions} concludes the paper.

\section{Optimal Perturbations}
\label{sec:NLOP}
Consider the finite-dimensional dynamical system
\begin{equation}
\dot{x} = f(x;X_0)
\end{equation}
where $x=x(t)\in\mathbb{R}^n$ is a perturbation away from a steady-state attractor $X_0$ of the nonlinear governing equations.
The nonlinear optimal perturbation~(NLOP) is defined as the solution to~\cite{kerswellAnnRev2018,kerswell2014}
\begin{maxi!}|l|[2]
    {x(0)}{\|x(T)\|^2_2}{}{}
    \label{eq:NLOP}
    \addConstraint{\dot{x}-f(x;X_0) = 0}
    \addConstraint{\|x(0)\|^2_2-d^2=0},
\end{maxi!}
where the initial perturbation energy $d^2$ and the length of the time-horizon $T$ are given. 
Note that defining the energy as $\|x(t)\|^2_2$ is without loss of generality.
Bisection over $d$ and $T$ can be used to determine the so-called ``minimal seed'' and the associated upper-bound for the perturbation energy $\|x(0)\|^2_2$ required to trigger instability away from the steady attractor $X_0$.

The equality constrained optimization problem can be converted to an unconstrained optimization problem by introducing the Lagrangian
\begin{equation}
\mathcal{L} = \|x(T)\|_2^2 + \int_0^T p\trans(t)\left[(\dot{x}(t)-f(x(t))\right] \mathrm{d}t+ \lambda(\|x(0)\|^2_2-d^2)
\label{eq:nloplag}
\end{equation}
where $\lambda\in\mathbb{R}$ and $p(t)\in\mathbb{R}^n$ are Lagrange multipliers.
The $p(t)$ is referred to as the \emph{co-state} or \emph{adjoint} in the optimal control literature.
Considering the first variation of the Lagrangian with respect to each of the variables yields the first-order necessary conditions for optimality:
\begin{subequations} \label{FirstO_conditions}
\begin{align}
\dot{x} &= f(x) \label{eq:perturb_subeq}\\
\dot{p} &= -\left(\frac{\partial f}{\partial x}\right)\trans p(t) \label{eq:adj_subeq}\\
0&=2\lambda x(0)-p(0)\\
0&=2x(T)+p(T)\\
0&=\|x(0)\|^2_2 - d^2. 
\end{align}
\end{subequations}
This system of two differential and three algebraic equations can be solved iteratively, for example, using gradient methods.
A basic implementation follows roughly as:

\begin{enumerate}
\item Initialize $x^{(0)}(0)$ to satisfy $\|x^{(0)}(0)\|^2_2=d^2$.
\item Given $x^{(i)}(0)$, integrate the primal system \eqref{eq:perturb_subeq} forward in time from $t=0$ to $t=T$ and store the solution $x^{(i)}(t)$. 
\item Given $x^{(i)}(t)$ and $p(T)=-2x(T)$, solve the co-state equation \eqref{eq:adj_subeq} backward in time from t=T to t=0.  Store $p^{i}(0)$.  
\item Evaluate the stopping criterion and terminate if $\left|\left|\frac{x^{(i)}(0)\trans p^{(i)}(0)}{d\|p^{(i)}(0)\|}\right| -1\right| <\epsilon$.  Otherwise, solve for $\lambda$ such that $\|x^{(i)}(0)+\Delta(2\lambda x^{(i)}(0)-p^{(i)}(0))\|^2_2=d^2$ and repeat from step 2 using $x^{(i+1)}(0)\leftarrow x^{(i)}(0) + \Delta(2\lambda x^{(i)}(0)-p^{(i)}(0)) $, where $\Delta>0$ is a parameter that defines the gradient step size.
\end{enumerate}
There are several practical issues that must be considered when implementing this direct adjoint looping method. The first is that the co-state $p(t)$ depends on the state $x(t)$; however, storing $x(t)$  over the full time horizon can be expensive.
Thus, a `check-pointing' procedure (see \cite{Berggren,Hinze2006AnOM}) can be implemented to reduce the storage requirements.
This is done by recalculating the (primal) state over short intervals during the backward integration of the co-state. This approach requires that $x(t)$ be stored at regular intervals $t=\tau_k$ during the forward time integration. 
The second practical consideration for gradient-ascent methods is selecting a suitable step size $\Delta$. To find a step size that yields sufficient accuracy for each iteration---while also balancing the rate of convergence---we perform an inexact line search using Armijo's rule \cite{Nocedal1999NumericalO}. We thereby ensure that the change in the optimization variable is proportional to the step length.

\subsection{Optimal Perturbations: The Linear Case}
Now consider the special case of linear time-invariant perturbation dynamics $\dot{x}(t)=Ax(0)$.
The state at time $T$ is related to the initial state as $x(T)=\Phi(T,0)x(0)$, where $\Phi(T,0)$ is the state transition matrix.
The transient amplification $G(T):=\|x(T)\|_2^2/\|x(0)\|_2^2$ will be invariant to the initial perturbation magnitude $d$ by virtue of linearity, and so we set it to unity.
Thus, for linear dynamics, the optimization in~\eqref{eq:NLOP} reduces to
\begin{maxi!}|l|[1]
    {\|x(0)\|^2_2=1}{\|\Phi(T,0)x(0)\|_2^2.}{}{}
    \label{eq:LOP}
\end{maxi!}
It is straightforward to show that the Lagrangian in this case is
\begin{equation}
\mathcal{L} = x\trans(0)P(T)x(0) - \lambda(x\trans(0)x(0)-1)
\end{equation}
where $P(T):=\Phi\trans(T,0)\Phi(T,0)>0$.
Setting the first variation of $\mathcal{L}$ with respect to $x(0)$ to zero yields
\begin{equation}
    \left(P(T) - \lambda I\right)x(0) = 0.
\end{equation}
This is a standard eigenvalue problem, with the optimal $x(0)$ corresponding to the eigenvector associated with the maximum eigenvalue $\lambda$ of $P(T)$.

\section{Sparse Optimal Perturbations}
\label{sec:spNLOP}
The standard NLOP problem can further be extended to investigate sparse optimal perturbations.
Sparse NLOP can be desirable when seeking, e.g.,~the single element in the admissible set of perturbations to concentrate the available perturbation energy $d^2$.
In principle, the sparse NLOP problem is combinatorially difficult.
Here, we exploit $\ell_1$-regularization to promote sparsity in the original NLOP problem:
\begin{maxi!}|l|[2]
    {x(0)}{\|x(T)\|_2^2 - \sigma\|x(0)\|_1}{}{}
    \addConstraint{\dot{x}-f(x;X_0) = 0}
    \addConstraint{\|x(0)\|_2^2-d^2=0}
\end{maxi!}
where again the initial perturbation energy $d^2$ and the length of the time-horizon $T$ are given.
Here, we have introduced a sparsity promoting parameter $\sigma\ge0$ into the objective function.
Solving the optimization problem for different values of $\sigma$ results in a family of optimal perturbations that define a Pareto front. When $\|x(0)\|_1$ is relatively small, the solution tends to be sparser than when $\|x(0)\|_1$ is relatively large.
Although $\ell_1$-regularization is a heuristic for sparsity, it allows for relatively efficient numerical procedures for solving the associated sparse NLOP problem.
When $\sigma=0$, we recover the standard NLOP problem/solution.
As $\sigma$ is increased, the solution will tend to become sparser (i.e.,~$x^*(0)$ will have fewer non-zero elements, or equivalently $x^*(0)$ will have a lower cardinality).

The Lagrangian for this problem follows as
\begin{eqnarray}
\mathcal{L} = &&\|x(T)\|_2^2 + \int_0^T p\trans(t)\left[\dot{x}(t)-f(x(t))\right]\mathrm{d}t\nonumber\\
&&+ \lambda(\|x(0)\|^2-d^2) -\sigma\|x(0)\|_1.
\end{eqnarray}
Note that this Lagrangian is simply the sum of the original Lagrangian~\eqref{eq:nloplag} from the standard NLOP problem, and a sparsity promoting term $-\sigma\|x(0)\|_1$.
This problem structure enables the use of an important class of solution algorithms known as \emph{proximal gradient methods}~\cite{beckBook}.
We will derive a convenient and easy-to-implement method based on the \emph{iterative soft thresholding algorithm~(ISTA)}, which is commonly used to solve optimization problems with a composite objective function that includes an $\ell_1$-regularization term \cite{beckBook}.
ISTA makes use of the soft threshold function, which is defined as
\begin{equation}
S_\sigma(y) := \begin{cases}y+\sigma & \text{ if } y<-\sigma\\ 0 & \text{ if } -\sigma\le y\le\sigma\\
   y-\sigma & \text{ if } y>\sigma. \end{cases}
\end{equation}
Thus, $S_\sigma$ serves to reduce the absolute value of the argument by no more than a given threshold on $\sigma\ge0$.
It follows that the sparse NLOP problem can be solved simply by solving a modified version of the standard NLOP problem---i.e.,~using the gradient ascent algorithm presented in Section~\ref{sec:NLOP} with a modified gradient calculation---with an application of $S_\sigma$ to each gradient step prior to advancing to the next iteratate.
Care must be taken with the gradient calculation, since the gradient of $\|\cdot\|_1$ consists of signum functions and is not continuously differentiable.
The sub-gradient at $\mathrm{sign}(0)$ is defined as the interval $[-1,1]$, and algorithms can be developed accordingly.
For simplicity here, we choose to remove the singularity at zero in the signum function by defining a regularized gradient
\begin{eqnarray}
    \nabla_\epsilon\|y\|_1= && W_\epsilon y, \nonumber\\
    \qquad W_\epsilon:=&& \mathrm{diag}(w_1,\dots,w_n), \nonumber\\
    \quad w_i:= && \begin{cases}1/|y_i| & \text{ if } |y_i|\ge\epsilon\\0 & \text{ if } |y_i|<\epsilon.\end{cases}
    \label{eq:regsignum}
\end{eqnarray}
A sparsity promoting DAL algorithm based on ISTA and a regularized gradient \eqref{eq:regsignum} roughly follows as:
\begin{enumerate}
\item Initialize $x^{(0)}(0)$ to satisfy $\|x^{(0)}(0)\|^2=d^2$.
\item Given $x^{(i)}(0)$, integrate the primal system \eqref{eq:perturb_subeq} forward in time from t=0 to t=T, storing $x^{(i)}(t)$ at $k$ predetermined checkpoints $t=\tau_k$. Store the solution $x^{(i)}(t)$. 
\item Given $x^{(i)}(t)$ and $p(T)=-2x(T)$, solve the co-state equation \eqref{eq:perturb_subeq} backward in time from t=T to t=0 utilizing the checkpoints $x^{(i)}(\tau_k)$.  Store $p^{i}(0)$.  
\item Evaluate the stopping criterion and terminate if $\left|\left|\frac{x^{(i)}(0)\trans p^{(i)}(0)}{d\|p^{(i)}(0)\|}\right| -1\right| <\epsilon$.  Otherwise, solve for $\lambda$ such that $\|x^{(i)}(0)+\Delta(2\lambda x^{(i)}(0)-p^{(i)}(0))\|^2=d^2$ and repeat from step 2 using\\ $x^{(i+1)}(0)\leftarrow$\\$S_\sigma\left(x^{(i)}(0) + \Delta(2\lambda x^{(i)}(0)-p^{(i)}(0)-\sigma W_\epsilon x^{(i)}(0)) \right)$.
\end{enumerate}
Just as for standard NLOP discussed in Section \ref{sec:NLOP}, we implement check-pointing for backward time integration of the co-state. Furthermore, an appropriate step size $\Delta$ is computed using an inexact line-search at each iterate.

\subsection{Sparse Optimal Perturbations: The Linear Case}
\label{sec:linopt_method}
For linear systems, several simplified solution procedures can be derived.
Specifically, consider the linear system $\dot{x}(t)=Ax(0)$.
The state at time $T$ is related to the initial state as $x(T)=\Phi(T,0)x(0)$.
It is straightforward to show that the Lagrangian in this case reduces to 
\begin{equation}
\mathcal{L} = x\trans(0)P(T)x(0) - \lambda(x\trans(0)x(0)-1) -\sigma\|x(0)\|_1
\end{equation}
where $P(T):=\Phi\trans(T,0)\Phi(T,0)>0$ and we have set $d=1$ as before.
Considering the variation with respect to $x(0)$ and regularizing the signum function as in \eqref{eq:regsignum}, the first-order necessary condition for optimality follows as
\begin{equation}
    (P(T) + \sigma W_\epsilon)x(0) = \lambda x(0).
    \label{eq:linopteigen}
\end{equation}
For $\sigma=0$, there will be no sparsity promotion, and we recover the standard eigenvalue problem associated with the optimal perturbation for maximizing the linear transient growth at time $t=T$: i.e.,
the optimal $x(0)$ will be the eigenvector associated with the maximum eigenvalue $\lambda$ determined from \eqref{eq:linopteigen}.
For $\sigma\ne0$, this is no longer a standard eigenvalue problem due to the fact that $W_\epsilon=W_\epsilon(x(0))$.
A very simple solution method is to iteratively solve a sequence of eigenvalue problems to obtain a sparse optimal perturbation $x(0)$.
That is, first solve~\eqref{eq:linopteigen} for the standard optimal perturbation $x(0)$, then use this $x(0)$ to determine $W_\epsilon(x(0))$.
This fixes $W_\epsilon$, and so now~\eqref{eq:linopteigen} can be solved as a standard (but modified) eigenvalue problem.
This process can be repeated until convergence. Alternatively, this problem can be solved using related methods used for solving sparse principal component analysis (PCA) problems~(see, e.g., \cite{journee2010}).

\section{Results}
\subsection{An Illustrative Example} \label{sec:AnExample}
Consider the nonlinear dynamics of perturbations for a two-dimensional system given by
\begin{equation} \label{2state_genform}
    \dot{x} = f(x;R)=A(R)x + Q(x)x
\end{equation}
where
\begin{eqnarray}
A(R)=&&\begin{bmatrix}-1/R&1\\0&-1/R\end{bmatrix},\nonumber\\
Q(x) =&& -Q\trans(x) = \begin{bmatrix}0 &-x_1\\ x_1 & 0\end{bmatrix}.  
\end{eqnarray}
This simple model possesses many features of the incompressible Navier-Stokes equations:
the scalar parameter $R>0$ acts like the Reynolds number; the linear term is non-normal; and the nonlinear term $Q(x)x$ is quadratic and energy conserving.
For a simple two-dimensional system, sparse optimal perturbations can be determined with relative ease using a brute force search.
Nonetheless, this simple example is instructive with regards to the features of sparse NLOP solutions in comparison with standard NLOP solutions.
The example also serves to validate the algorithm, which is facilitated by visual inspection of phase portraits.

The energy threshold to instability for a sparse NLOP is found to be greater than that for a standard NLOP.
This is to be expected; in general, the energy threshold to instability for a sparse minimal seed will be no less than the energy threshold for a non-sparse minimal seed.
For this two dimensional example, a sparse non-zero solution will have a single non-zero element.
Here we will investigate the case with $R=2\sqrt{2}$, which is greater than the critical $R$ for transient growth in this system.
Applying bisection on $d$ with a tolerance of 0.01 and fixing $T=3.6$ and $\sigma=0.2d$, we find upper bounds on the disturbance threshold for instability: $d^*\le0.12$ for the NLOP and $d^*_{sp}\le0.13$ for the sparse NLOP.
We note that the NLOP and sparse NLOP problems can only provide upper bounds to instability thresholds.  
Any additional bisection over the other optimization parameters could only serve to provide lower values for these upper bounds.
Recently proposed convex-optimization-based methods can be used to obtain lower bounds on these instability thresholds for the non-sparse NLOP problem in quadratic systems~\cite{kalurPRF2021,kalurLCSS2022,tosoLCSS2022,liaoCDC2022}, which would also provide lower-bounds for the sparse NLOP problem.

\begin{figure}[h!]
\subcaptionbox{$d=0.1$}{\includegraphics[width=0.32\textwidth]{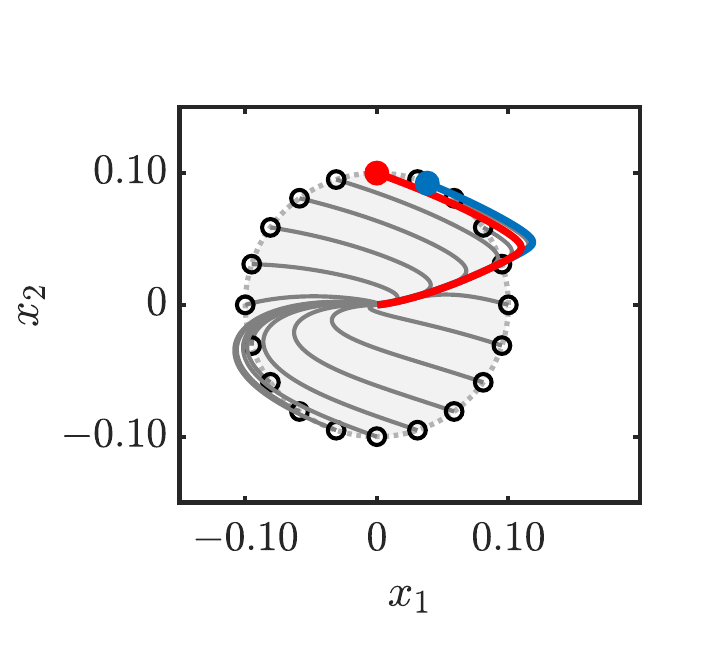}}
\subcaptionbox{$d=0.12$}{\includegraphics[width=0.32\textwidth]{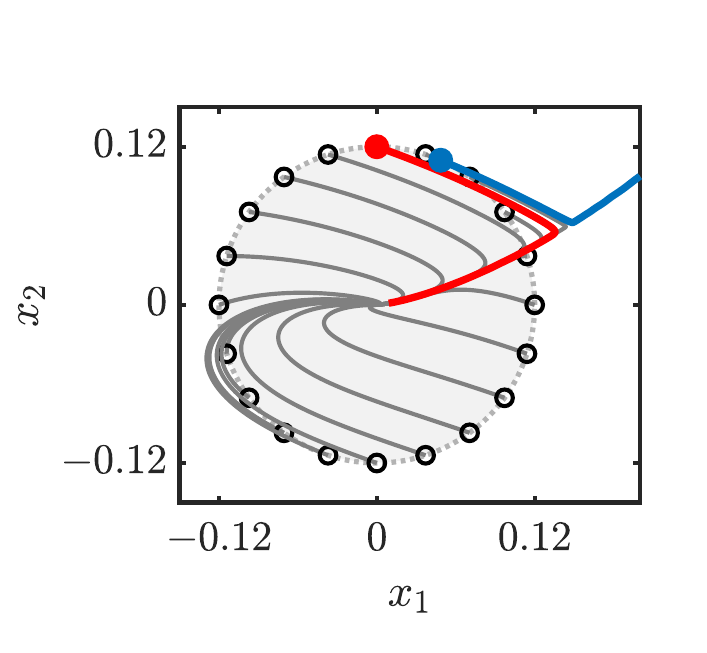}}
\subcaptionbox{$d=0.13$}{\includegraphics[width=0.32\textwidth]{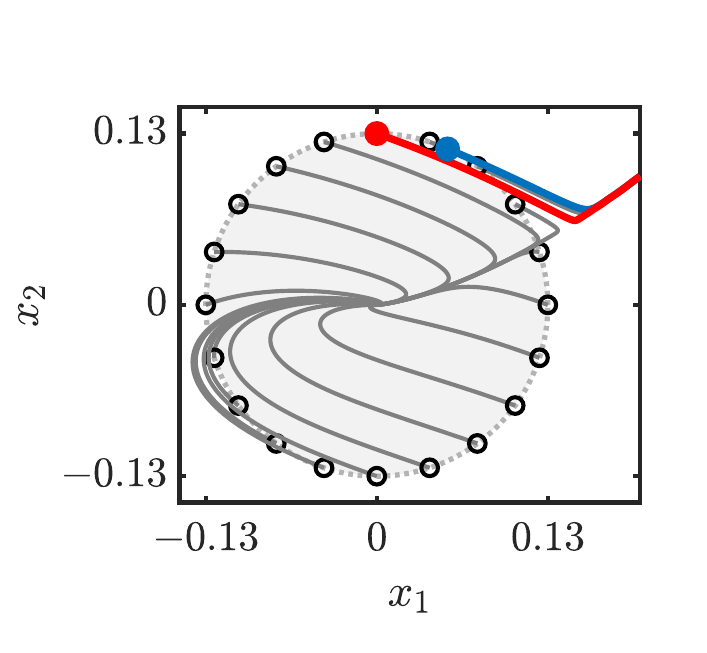}}
\caption{The nonlinear optimal perturbation (blue) drives the state further from the steady attractor than the sparse nonlinear optimal perturbation (red) as shown in (a). The NLOP triggers instability at a lower threshold disturbance than the sparse NLOP as shown in (b) and (c). Nonlinear system responses for a set of sub-optimal initial perturbations are also plotted in gray. The light gray circle indicates the region for which energy is less than~$d^2$.}
\label{fig:spNLOP}
\end{figure}

The system response for three disturbance levels are reported in Figure~\ref{fig:spNLOP}.
For $d=0.10$, the NLOP $x^*$ and the sparse NLOP $x^*_{sp}$ solutions computed are
\begin{equation*}
x^*=\begin{bmatrix} 0.0384&
0.0923\end{bmatrix}\trans, \qquad x^*_{sp}=\begin{bmatrix}0&
    0.1\end{bmatrix}\trans.
\end{equation*}
%
For $d=0.12$, we find
\begin{equation*}
x^*=\begin{bmatrix} 0.0485&
0.1098\end{bmatrix}\trans, \qquad x^*_{sp}=\begin{bmatrix}0&
    0.12\end{bmatrix}\trans
\end{equation*}
and for $d=0.13$ we find
\begin{equation*}
x^*=\begin{bmatrix} 0.0538&
0.1183\end{bmatrix}\trans, \qquad x^*_{sp}=\begin{bmatrix}0&
    0.13\end{bmatrix}\trans.
\end{equation*}
These reported solutions all constitute global optima.  Local optima can be found by initializing the associated DAL algorithms differently. A simple heuristic for doing so based on LOP analysis is described below.
\onecolumngrid

\twocolumngrid

\begin{figure}[h!]
\centering
\includegraphics[width=0.32\textwidth]{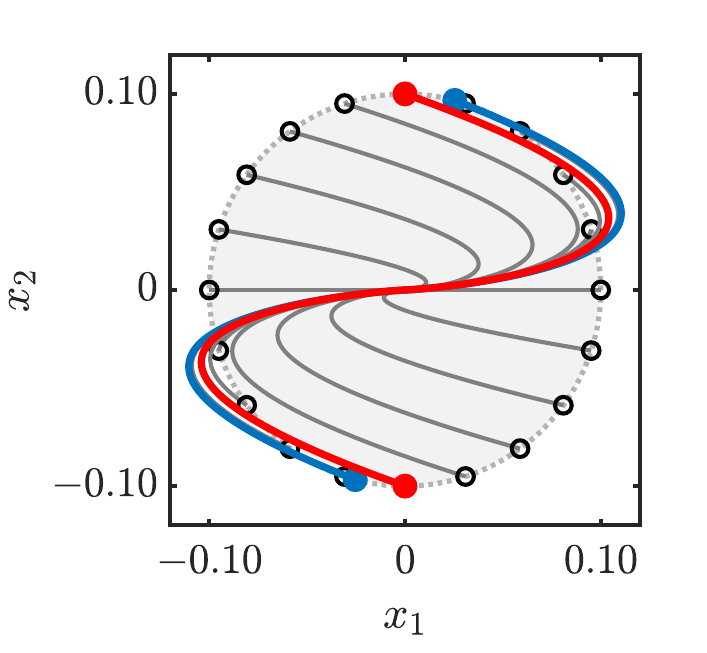}
\caption{Linear optimal perturbations (blue) drive the state further from the steady attractor than sparse linear optimal perturbations (red).  Linear system responses for a set of sub-optimal initial perturbations are also plotted in gray.}
\label{fig:splinop}
\end{figure}

Next, we consider the associated linear dynamics about the stable attractor.
Figure~\ref{fig:splinop} highlights the responses due to a linear optimal perturbation and a sparse linear optimal perturbation found by applying the method presented in Section~\ref{sec:linopt_method}:
\begin{equation*}
x^*=\begin{bmatrix}    0.4472&0.8944\end{bmatrix}\trans, \qquad x^*_{sp}=\begin{bmatrix}0&
    1\end{bmatrix}\trans.
\end{equation*}
Of note in this example is that the sparse linear optimal perturbation identified coincides with the sparse NLOP identified previously.
Of course, for a linear system, the (sparse) optimal solution will never be unique---even when a unique optimal eigendirection is identified: if $x(0)=x^*$ is optimal, then so is $x(0)=-x^*$.
Nonetheless, this provides a heuristic for efficiently finding sparse NLOP solutions:
First, solve for all sparse linear optimal perturbations $\{x_i^*\}$, then seed the nonlinear simulation with $\{d\cdot x_i^*\}$, and find the trajectory with $\|x_i(T)\|$ being the maximum.
In most cases, this will require two simulations for a given perturbation amplitude $d$ and time-horizon $T$.
This heuristic can guide initialization of DAL algorithms to isolate candidates for the globally optimal solution to the (sparse) NLOP problem.
\subsection{Reduced-Order Model of a Sinusoidal Shear Flow} \label{sect:model_overview}
Next, we demonstrate the sparse NLOP framework on a reduced-order model of a shear flow wherein the fluid experiences a sinusoidal body force between two free-slip moving walls \cite{Moehlis2004}. This model has the form

\begin{equation} \label{9-state form}
\dot{x} = f(x;Re) = A(Re)x+ Q(x)x,
\end{equation}
where the state vector $x=x(t)\in\mathbb{R}^{9}$.
The linear term $A(Re) \in \mathbb{R}^{9\times9}$ is Hurwitz, and is parameterized by the Reynolds number ${Re}>0$. The nonlinear term $Q(x)x$ in equation \eqref{9-state form} is a quadratic function
\begin{equation} 
    Q(x)x=
    \begin{bmatrix}
        x^TQ^{(1)}x \\
        \vdots \\
        x^TQ^{(9)}x
    \end{bmatrix},
\end{equation}
where $Q^{(1)},...,Q^{(9)}\in\mathbb{R}^{9\times 9}$ are symmetric matrices defined such that the quadratic nonlinearity is lossless.

The 9-state sinusoidal shear flow model presented here is a generalization of the well-established 8-state Waleffe model \cite{Waleffe} and has been used in a number of previous works to demonstrate novel stability analysis methods (see, e.g.,~\cite{Goulart2012,Liu2020,Kalur2022,Mushtaq2022}). The model also benefits from being physically interpretable as each of the nine Fourier modes describes a well-known feature that contributes to the dynamics of turbulence in many shear flows. The introduction of the ninth mode---which  describes the turbulence-induced modification of the basic flow profile---is the distinguishing feature of this 9-state model over the 8-state Waleffe model~\cite{Moehlis2004}.
Note that the dynamics in~\eqref{9-state form} possess a fixed-point at the origin.
This was achieved by shifting the state vector $\tilde{x}\in\mathbb{R}^9$ from the original dynamical equations in~\cite{Moehlis2004} for which the fixed-point is located at $\tilde{x}_1=1,\tilde{x}_2=...=\tilde{x}_9=0$ for all $Re$.
The coordinate shift is accounted for in the  linear term as $A(Re)=\tilde{A}(Re)+W$, where $\tilde{A}(Re)$ is the linear term from~\cite{Moehlis2004}, and $W\in \mathbb{R}^{9\times9}$ is defined such that for $c=[1,0,...,0]\trans\in\mathbb{R}^{9}$:
\begin{equation}
    W\tilde{x}=Q(\tilde{x})c+Q(c)\tilde{x}.
\end{equation}
The state vector $\tilde{x}$ contains modal coefficients that provide physical interpretability of the dynamics and allow for reconstruction of the velocity field $\boldsymbol{u}(\boldsymbol{z},t)$ as,

\begin{equation} \label{GalerkinVel}
    \boldsymbol{u}(\boldsymbol{z},t)=\sum_{m=1}^{9}{\tilde{x}_m}(t)\boldsymbol{\phi}_m(\boldsymbol{z}),
\end{equation}
where $\boldsymbol{\phi}_m(\boldsymbol{z})$ are the velocity modes and $\tilde{x}_m(t)$ are the associated modal coefficients. 
An overview of each mode and its physical relevance is provided in Table \ref{tab:MoehlistModes}. For more details on the model, the reader is referred to the original source in \cite{Moehlis2004}.

\begin{table*}
\caption{\label{tab:MoehlistModes}The modes described by the 9-state model \cite{Moehlis2004} are physically interpretable and carefully constructed to model observed dynamics in turbulent flow simulations.}
\begin{ruledtabular}
\begin{tabular}{c l l r}
\textbf{Mode}       & \textbf{Name}       & \textbf{Description} \\ \hline
    $\tilde{x}_1$      & basic profile   & Describes the mean velocity profile of the flow.     \\ \hline
    $\tilde{x}_2$      & streak    & Captures spanwise variation of the streamwise velocity.  \\ \hline
    $\tilde{x}_3$      & downstream vortex  & Describes streamwise vorticies spanning the entire gap. \\ \hline
    $\tilde{x}_4,\tilde{x}_5$      & spanwise flow modes       & Sinusoidal streak-flow instabilities causing velocity perturbations of the streaky flow.  \\ \hline
    $\tilde{x}_6,\tilde{x}_7$   & normal vortex modes       & Generated by the advection of $\tilde{x}_4,\tilde{x}_5$ by the streak $(\tilde{x}_2)$ and the vortex $(\tilde{x}_3)$ modes.  \\ \hline
    $\tilde{x}_8$      & 3D mode       & Fully three-dimensional mode created by the interaction of modes $\tilde{x}_2$--$\tilde{x}_7$.  \\ \hline
    $\tilde{x}_9$      & base flow modifier       & Modification to the structure of the basic velocity profile ($\tilde{x}_1$) by the turbulence. \\ 
\end{tabular}
\end{ruledtabular}
\end{table*}

\begin{figure*}[h!]
\centering
\centering
\begin{subfigure}{0.45\textwidth}
    \includegraphics[width=\textwidth]{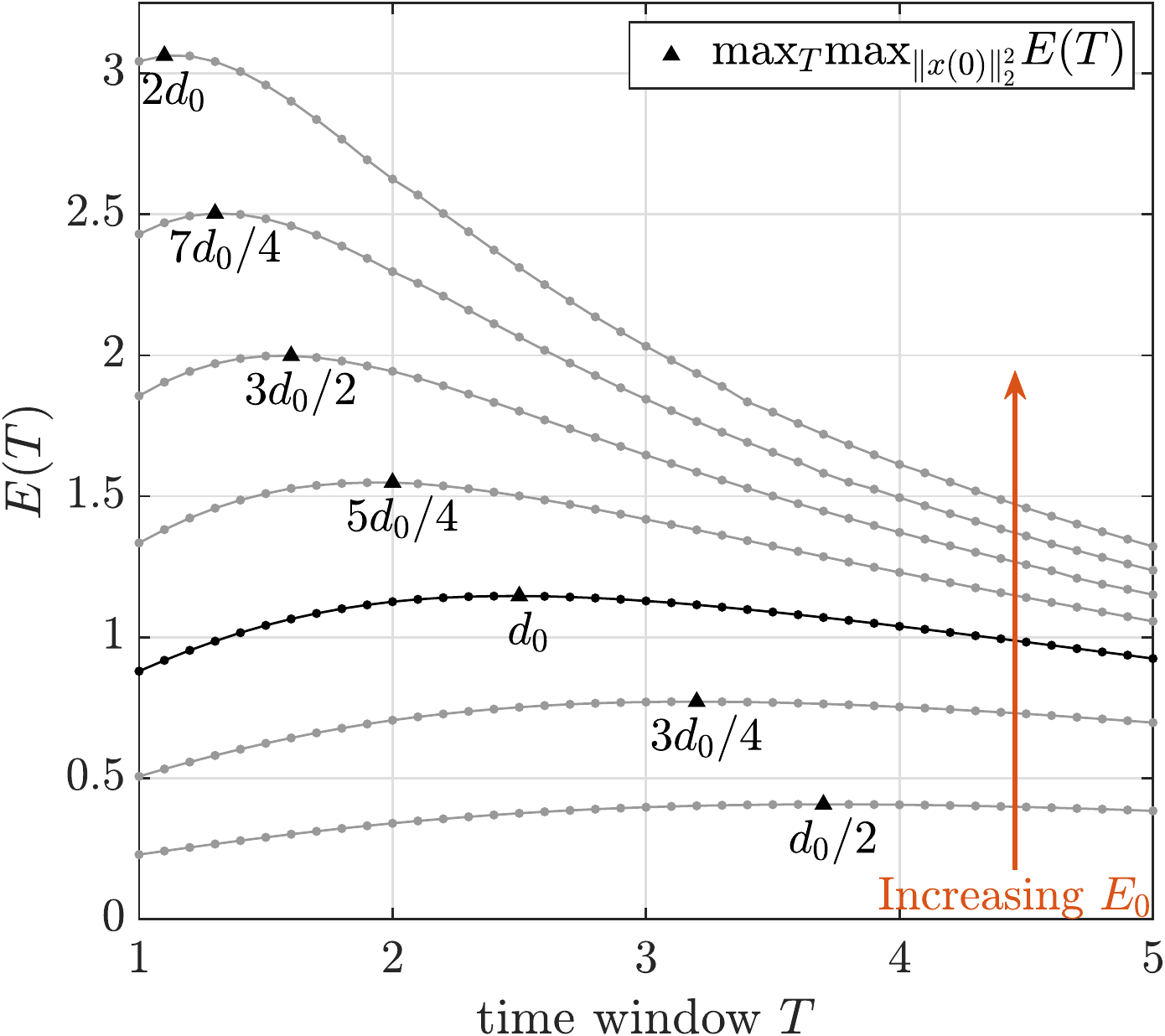}
    \caption{$E(T)$}
    \label{fig:EvsDa}
\end{subfigure}
\begin{subfigure}{0.45\textwidth}
    \includegraphics[width=\textwidth]{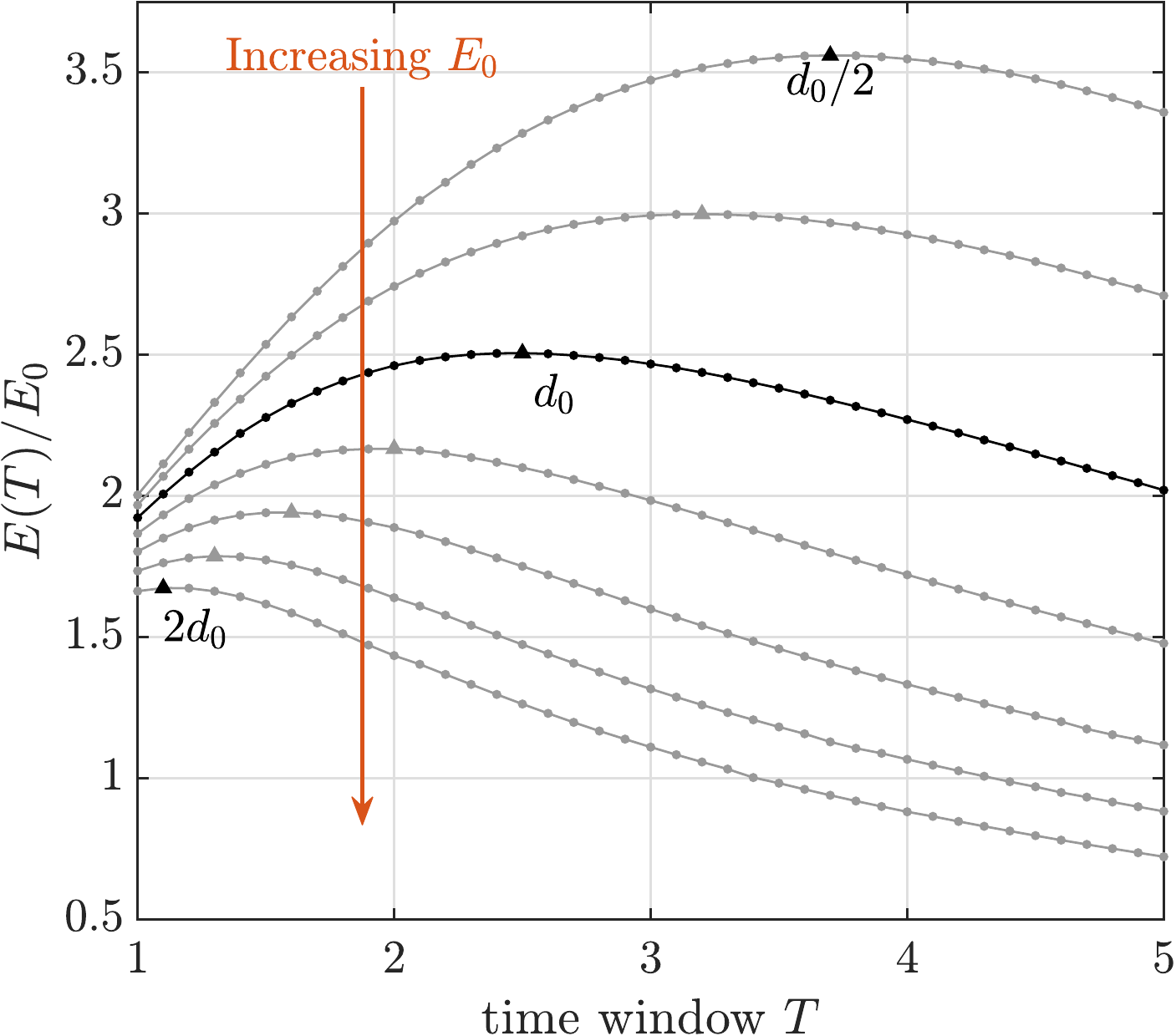}
    \caption{$E(T)/E_0$}
    \label{fig:EvsDb}
\end{subfigure}
\caption{Increasing the initial perturbation size $E_0$ pushes the the energy-maximizing flow state closer to $t=0$, resulting in a quicker return to energy decay $E(T)/E_0<1$. The sparsity parameter $\sigma$ was tuned so that the cardinality of the initial perturbations would be $k=4$. As shown in Figure \ref{fig:EvSP}, the energies are approximately the same for cardinalities between $k=4$ and $k=9$, so we chose to show $k=4$ as the sparsest of these cases.}
\label{fig:EvsD}
\end{figure*}

We now demonstrate the sparse NLOP method for $Re=20$.  
In order to focus our investigation on the transient growth phenomenon, we set a perturbation magnitude $d$ for which perturbations are guaranteed to remain within the region of attraction of the steady baseflow.
We employ the quadratic-constraint-based local stability analysis method described in \cite{Kalur2022} to estimate a bound on the permissible perturbation amplitude.
This quadratic constraint approach indicates that a perturbation amplitude of $d=d_0:=0.676$ is certifiably below the threshold for baseflow instability.
Using $d_0$, we can perform sparse NLOP to find an optimal perturbation that maximizes kinetic energy at time $T$, where the kinetic energy is defined as
\begin{equation} \label{eq:basic_energy}
    E(T)=||x(T)||_2^2.
\end{equation}
In addition to the initial perturbation kinetic energy $E(0)=E_0=d^2_0$, we also need to define the length of the time-horizon $T$. 
The black curve in Figure~\ref{fig:EvsDa} shows the maximum $E(T)$ obtained by performing sparse NLOP over a grid of forty $T$ values uniformaly spaced between $T=1$ and $T=5$. For brevity, the results reported in Figure~\ref{fig:EvsDa} correspond to a single sparsity promoting parameter $\sigma$ for which the cardinality of the optimal perturbations was $k=4$. The curves obtained for $k=9$ through $k=4$ were nearly identical, while those for $k=3$ and $k=1$ had less amplification $E(T)$ but followed the same trend.
For $d=d_0$, the maximum $E(T)$ over all $T$ is found to correspond to a time horizon of $T=2.5$.
For this case, the maximized energy $E(T=2.5)=1.28$ was the largest.
To examine the transient growth, we consider the amplification of the initial perturbation as,
\begin{equation} \label{eq:TEG}
    \frac{E(T)}{E_0}=\frac{||x(T)||_2^2}{d^2}.
\end{equation}
The black curve in Figure~\ref{fig:EvsDb} shows that for $d=d_0$ and $T=2.5$, the initial perturbation energy, $E_0$, is amplified by a factor of approximately 2.5.

As is demonstrated by the gray lines in Figure \ref{fig:EvsD}, increasing the perturbation magnitude $d$ results in an earlier and larger maximum energy $E(T)$. In other words, the identified initial perturbations $x^*$ tend to drive the flow away from its steady attractor over a shorter time as $E_0$ is increased. However, because the amplification is maximized at an earlier time, increasing the initial energy $E_0$ causes a decrease in the maximum transient energy growth. In other words, the identified initial perturbations become less efficient at driving the flow away from the steady attractor, due to a higher input energy being required.

\begin{figure*}[h!]
\centering
\includegraphics[width=.85\textwidth]{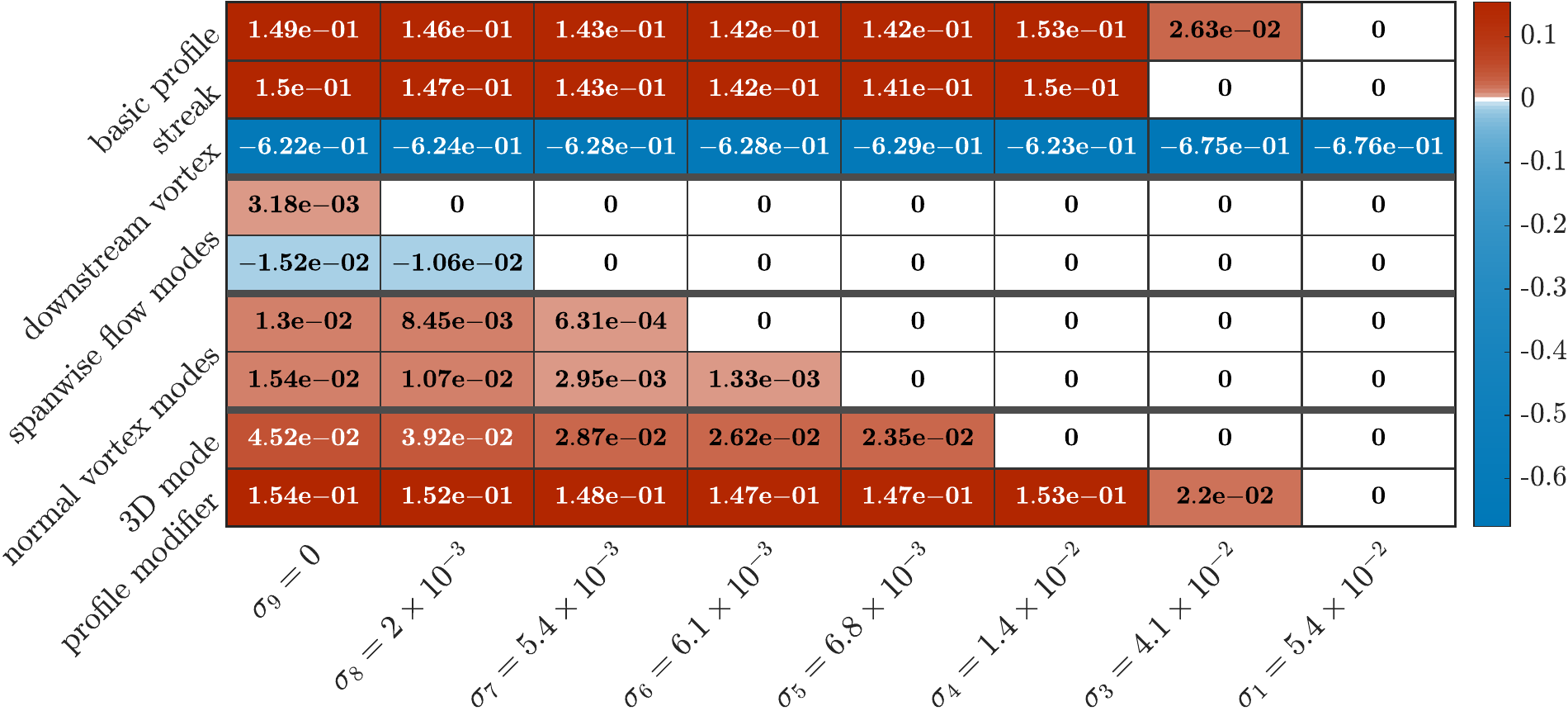}
\caption{Sparsity patterns in the optimal perturbations found for the case where  $Re=20$, $T=2.5$, and $d_0\approx0.676$. The columns each correspond to a sparsity parameter $\sigma$ that maximized energy for the particular sparsity value, and the rows correspond to the Fourier modes. Note that every optimal perturbation satisfies $||x^*||^2=d^2$.}
\label{fig:Sparsity}
\end{figure*}

\begin{figure*}[h!]
\centering
\includegraphics[width=0.80\textwidth]{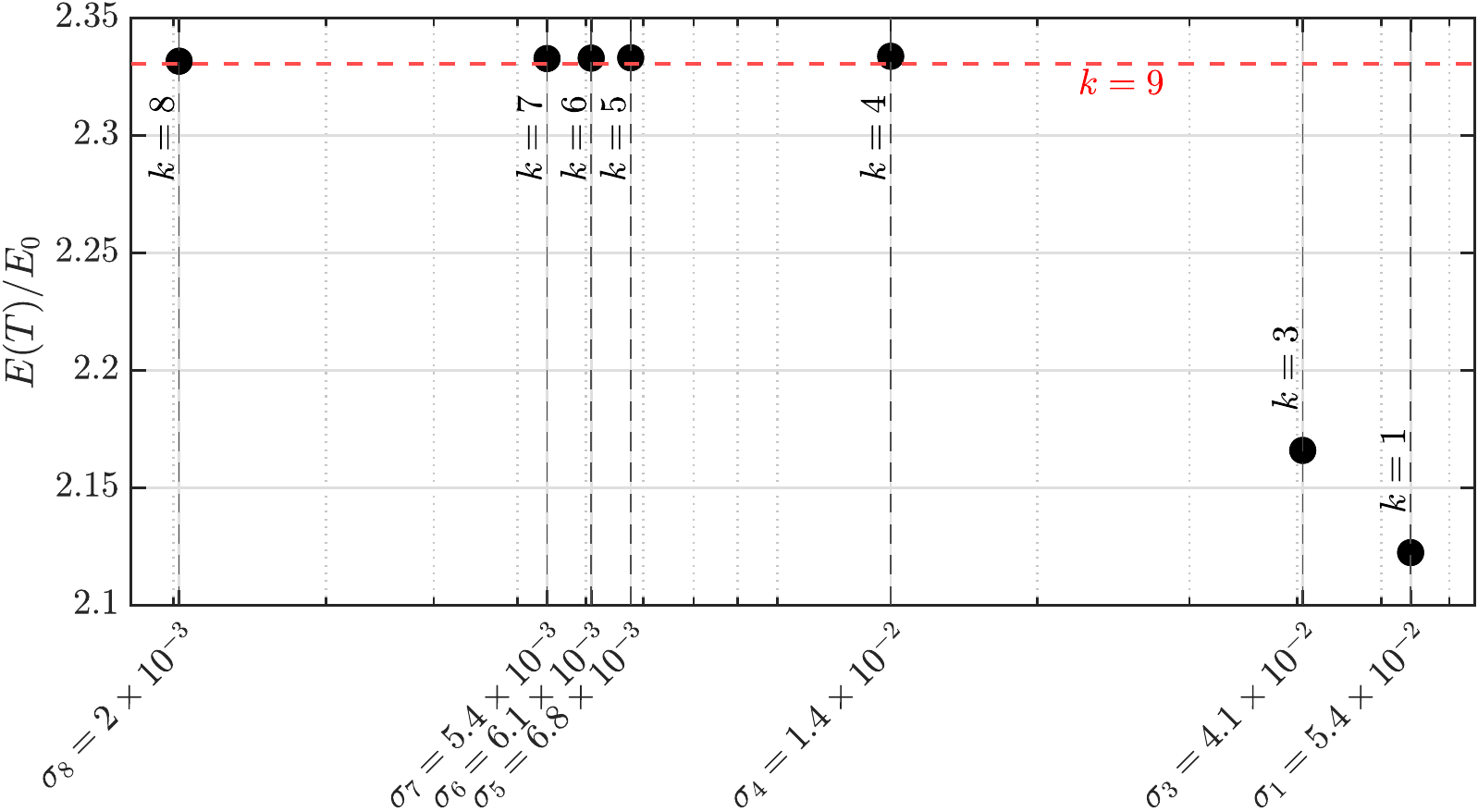}
\caption{The transient energy growths corresponding to each sparsified models shown in Figure \ref{fig:Sparsity} vs. the sparsification variable $\sigma$ for $T=2.5$. }
\label{fig:EvSP}
\end{figure*}

The sparsity promoting framework described in Section \ref{sec:spNLOP} has no guarantee of producing a specific cardinality $k$. As such, it is necessary to search over a range of $\sigma$ values to obtain optimal perturbations with different cardinalities. To do this, $d$ and $T$ must first be fixed as constants. We report results for the perturbation size $d=d_0=0.676$, which was shown in Figure~\ref{fig:EvsD} to correspond to a time horizon of $T=2.5$. As discussed in Section \ref{sec:AnExample}, a linear estimate is used to seed the non-sparse NLOP, which gives the optimal perturbation vector $x^*$ corresponding to the column labeled ($\sigma_9=0$) in Figure \ref{fig:Sparsity} (the subscript on $\sigma$ indicates the cardinality $k$). The resultant solution is used as a seed to perform sparse NLOP for a range of $\sigma$ values, revealing optimal perturbations as shown in Figure~\ref{fig:Sparsity}. Note that we do not report a result for $k=2$, as we were unable to find a corresponding $\sigma$ value using our search grid. The energy amplification associated with each sparse optimal perturbation presented in Figure~\ref{fig:Sparsity} is shown in Figure~\ref{fig:EvSP}. Note that the $\sigma$ axis is on a log scale, so the energy amplification for $\sigma_9=0$ is reported as the dashed line. 
The energies at $T=2.5$ associated with the optimal perturbations for  $\sigma_9$ through $\sigma_4$ do not differ significantly, while the two sparsest models correspond to significantly smaller maximum energies at time $T=2.5$. 

The convergence properties of gradient-based optimization are a well known weakness of this class of methods~\cite{beckBook}, and are likely the cause of the $0.1\%$ increase in transient energy between $k=9$ and $k=4$.
Thus, the comparable transient energy growth between the non-sparse optimal perturbation ($k=9$) and the sparse optimal perturbation with $k=4$ suggests a strong sensitivity to the initial seed and algorithmic constants such as the convergence tolerance $\epsilon$. 
By seeding the non-sparse NLOP with a linear heuristic, more gradient ascent steps were needed to converge to tolerance $\epsilon$ when compared to sparse NLOP. We note that the  method is also highly sensitive to the gradient step size $\Delta$, which is why we used an inexact line search to determine $\Delta$.
We also note that although the energies shown in Figure \ref{fig:Sparsity} are maximized, the sparsification values may differ depending on the initial guess for $x^*$.

\begin{figure*}[h!]
\centering
\includegraphics[width=0.80\textwidth]{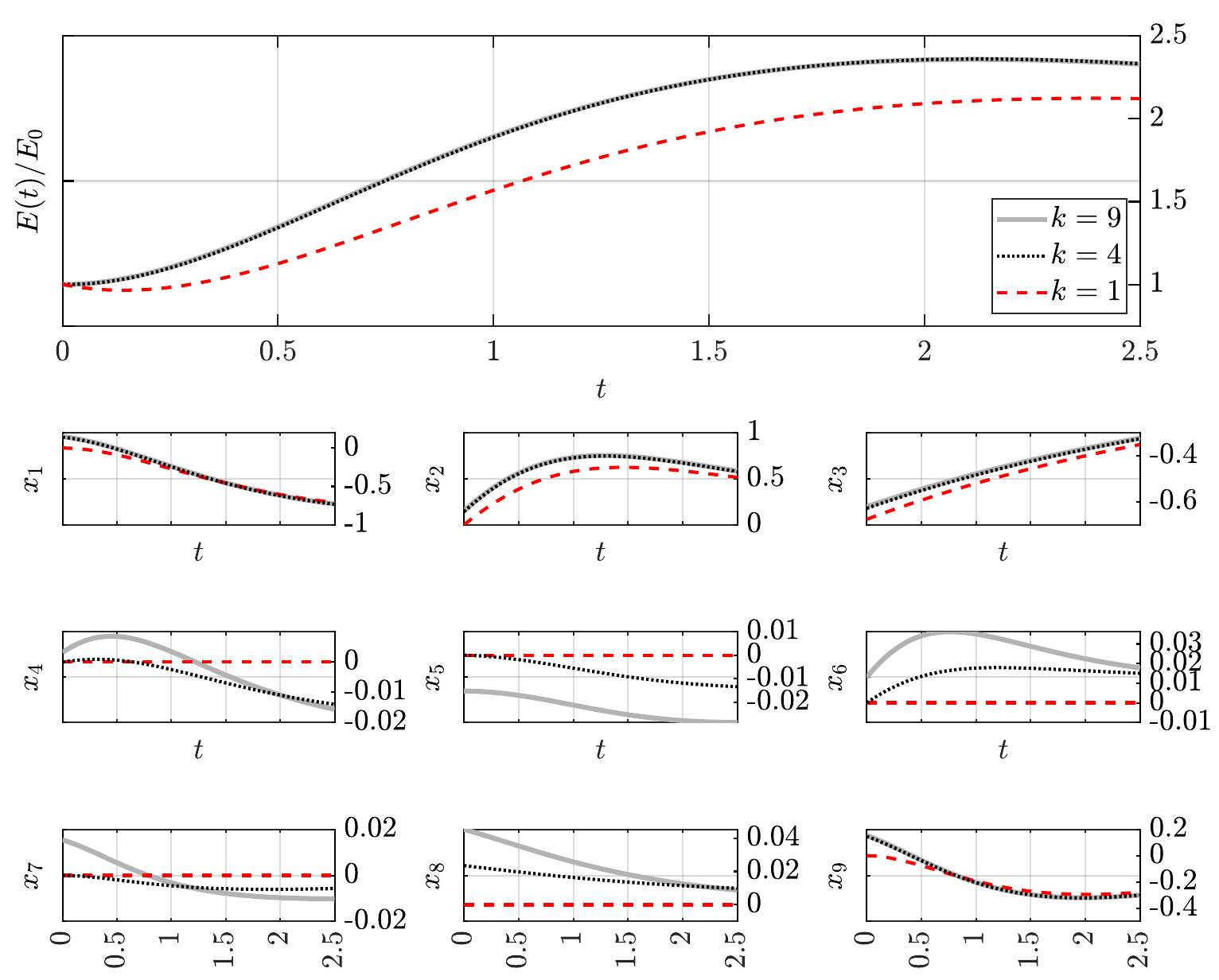}
\caption{Energy and mode-trajectories for the models obtained from $\sigma_9$, $\sigma_4$, and $\sigma_1$ in Figure \ref{fig:Sparsity}. The energy $E(T=2.5)/E_0$ for $k=1$ is lower than for the non-sparse cases, as modes $x_4$ through $x_8$ remain at zero for all $t$.}
\label{fig:k_traj}
\end{figure*}

The trajectories of the individual modes and the associated transient energies are shown in Figure \ref{fig:k_traj} for cardinalities $k=9$, $k=4$, and $k=1$. The case when $k=4$ is noteworthy, as this is the sparsest result that causes an amplification of all nine modes. We note that the spanwise flow modes $(\tilde{x}_4,\tilde{x}_5)$ are pruned first, followed by the normal vortex modes $(\tilde{x}_6,\tilde{x}_7)$. In the case of $Re=20$, the modes $\tilde{x}_4$--$\tilde{x}_7$ are not as dynamically significant as the others, which is reflected in their amplification being an order of magnitude lower than the other modes. The 3D mode $(\tilde{x}_8)$ is pruned next for the same reason.
This results in $x_4$--$x_8$ experiencing no amplification when $k=1$. 

The sparsity patterns shown in Figure \ref{fig:Sparsity} reveal that actuation of the vortex mode $x_3$ is central to maximizing transient energy growth. Furthermore, the sparsity pattern highlights the dominance of modes $x_1$ and $x_9$. This result is particularly noteworthy because the authors of the 9-state model emphasize the importance of the basic profile modification mode $x_9$ in the formulation of the model \cite{Moehlis2004};
the sparse NLOP identifies $x_9$ as a dynamically significant mode in the model without any prior knowledge of this fact.  We note that the total transient energy growth corresponding to $k=4$, as well as the trajectories of modes $x_1$--$x_3$ and $x_9$ are effectively the same for $k=4$ and the non-sparse $k=9$ case. These observations provide insight into how the modes are coupled, and which modes are dominant in driving instability.

 We will now show that the sparse optimal perturbations identified lead to  mechanisms that are similar to those observed during a transition to turbulence. Using Equation~\eqref{GalerkinVel} and the optimal initial perturbations, we first obtain velocity fields for the sinusoidal shear flow. We observe that for $Re=20$, the flow exhibits similar features for all initial perturbations shown in Figure~\ref{fig:EvsD}, regardless of cardinality. Furthermore, the downstream vortex mode $x_3$ is maximized for all aforementioned optimal perturbations. A comparison of flow states corresponding to two disturbance sizes and cardinalites is provided in Figure~\ref{fig:Flow_nonspVssp}.

\begin{figure*}[h!]
\subcaptionbox{For $k=1$, modes $x_4$--$x_8$ are zero for all $t$.}{\includegraphics[width=0.45\textwidth]{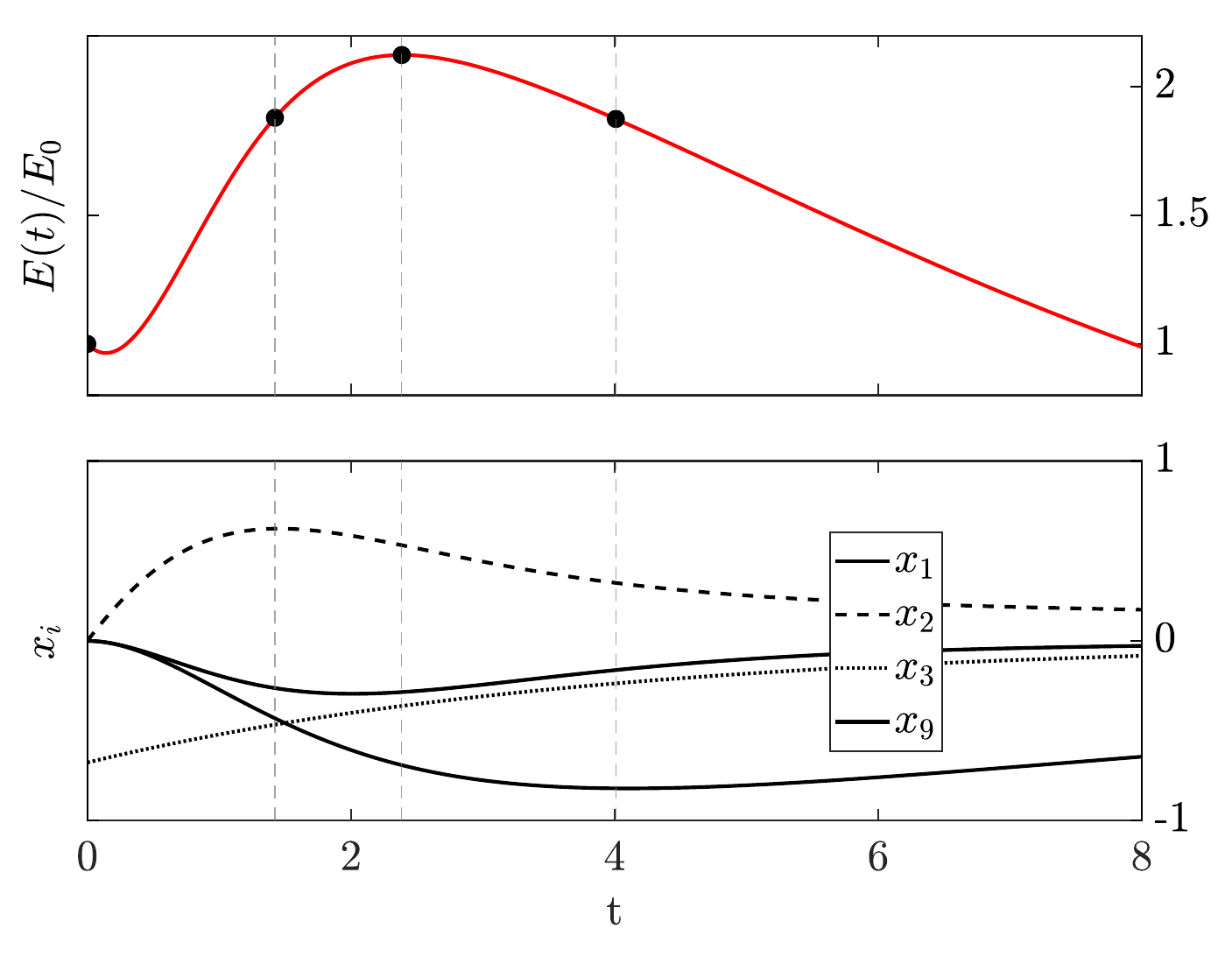}}
\subcaptionbox{$t=0$: Streak formation through initial vortices. }{\includegraphics[width=0.45\textwidth]{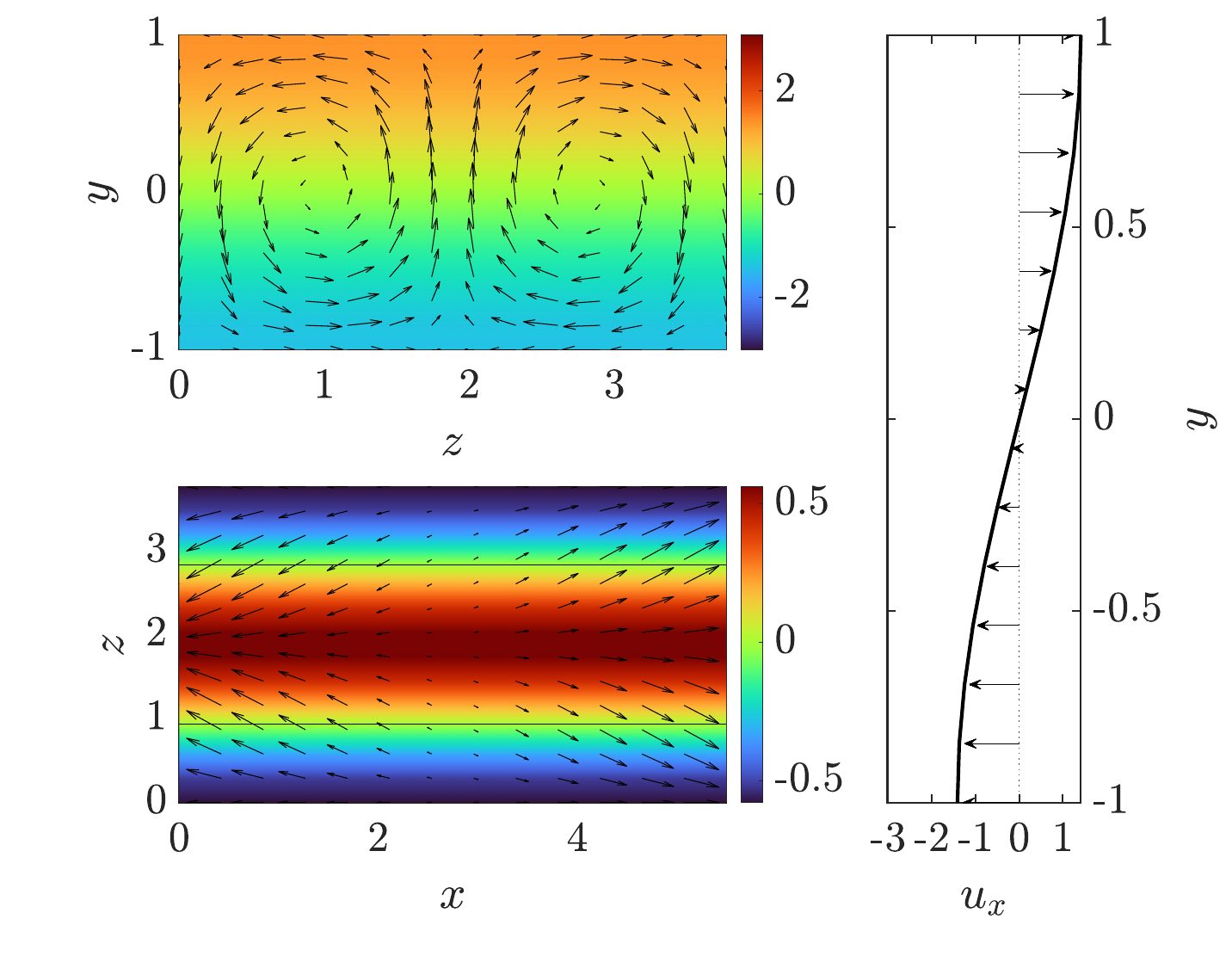}}
\subcaptionbox{$t=1.4$: The streak mode ($x_2$) is maximized.}{\includegraphics[width=0.45\textwidth]{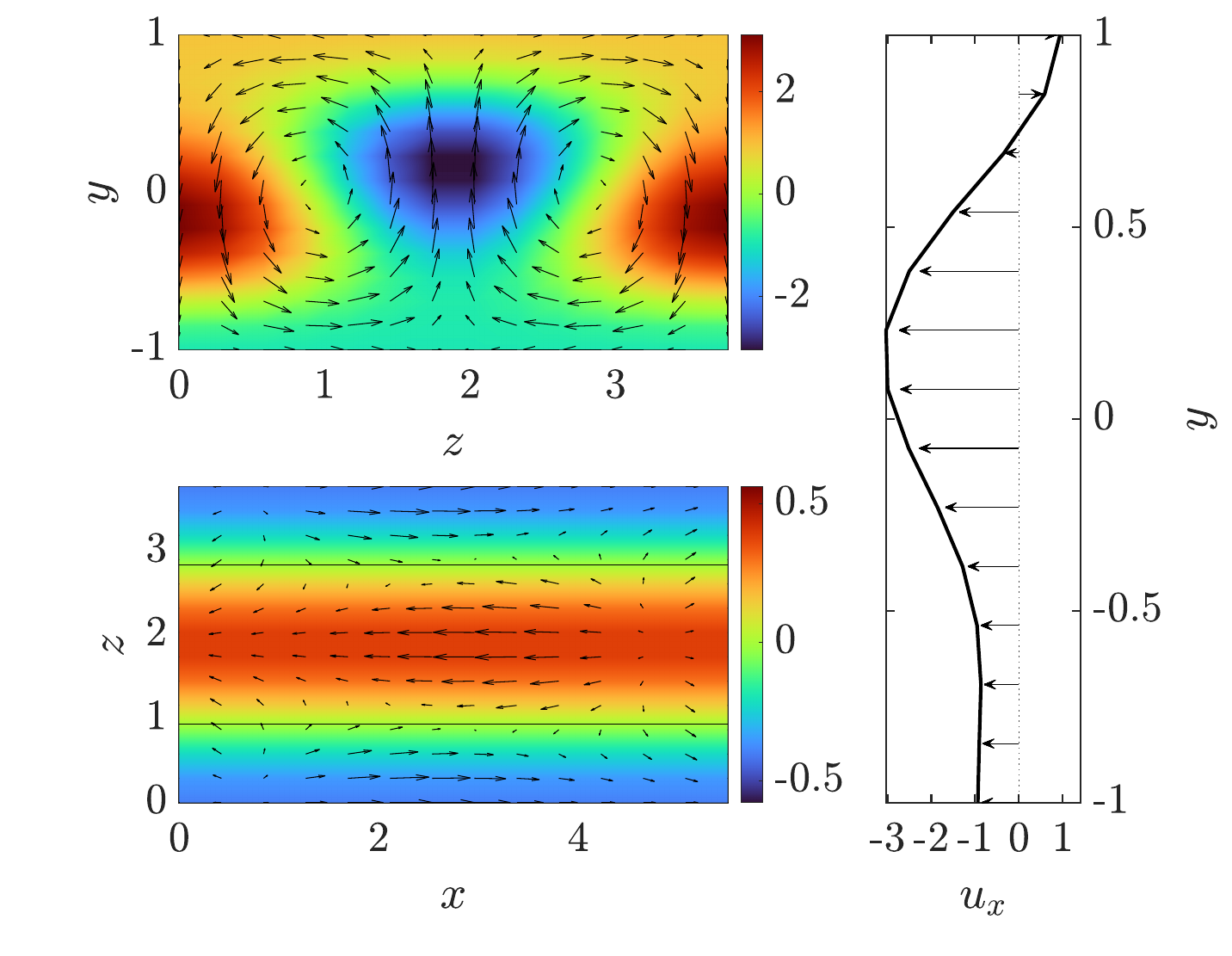}}
\subcaptionbox{$t=2.5$: Transient energy $E(t)/E_0$ is maximized.}{\includegraphics[width=0.45\textwidth]{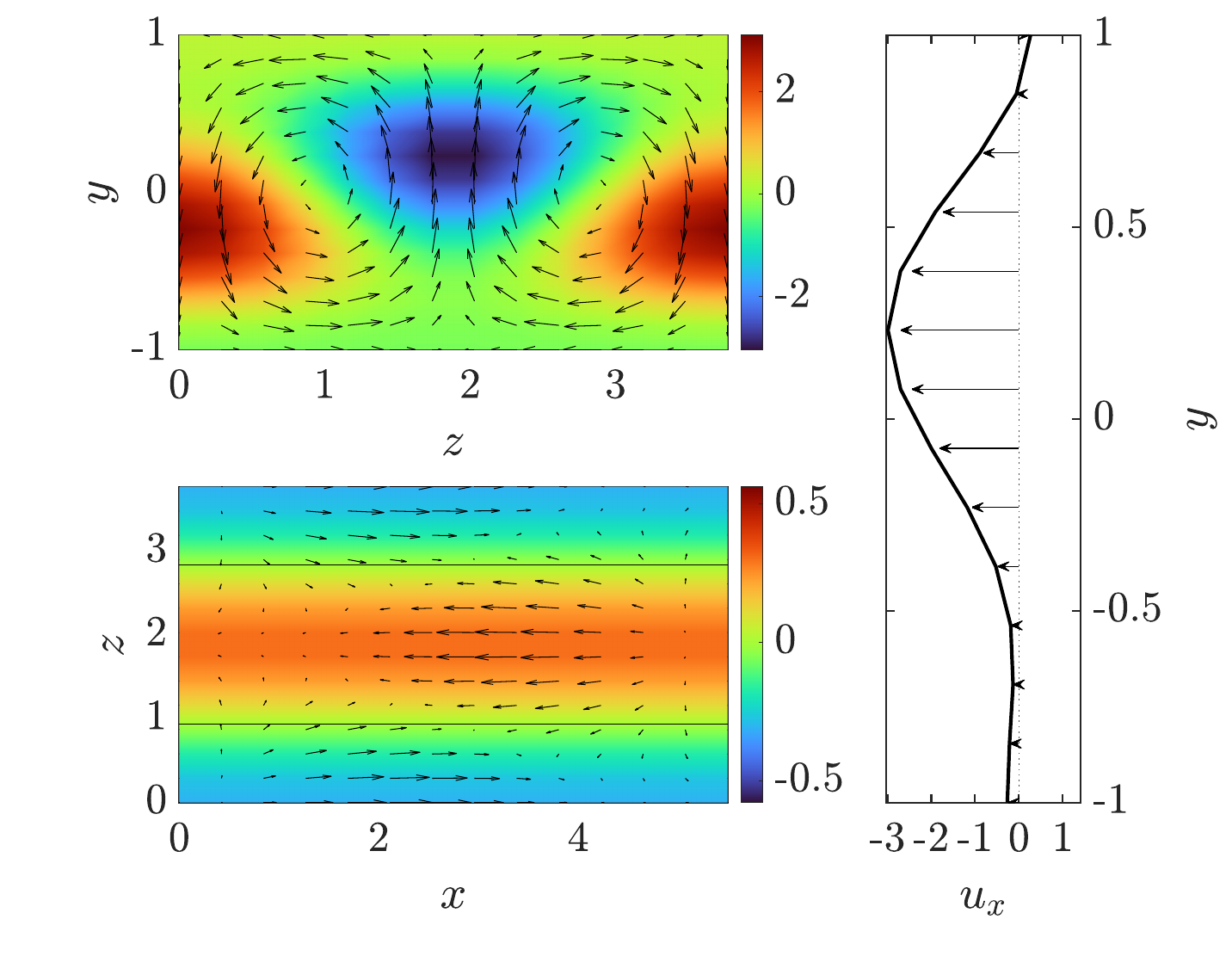}}
\subcaptionbox{$t=4$: Magnitude of $x_1$ maximized, the streaks begin to break down.}{\includegraphics[width=0.45\textwidth]{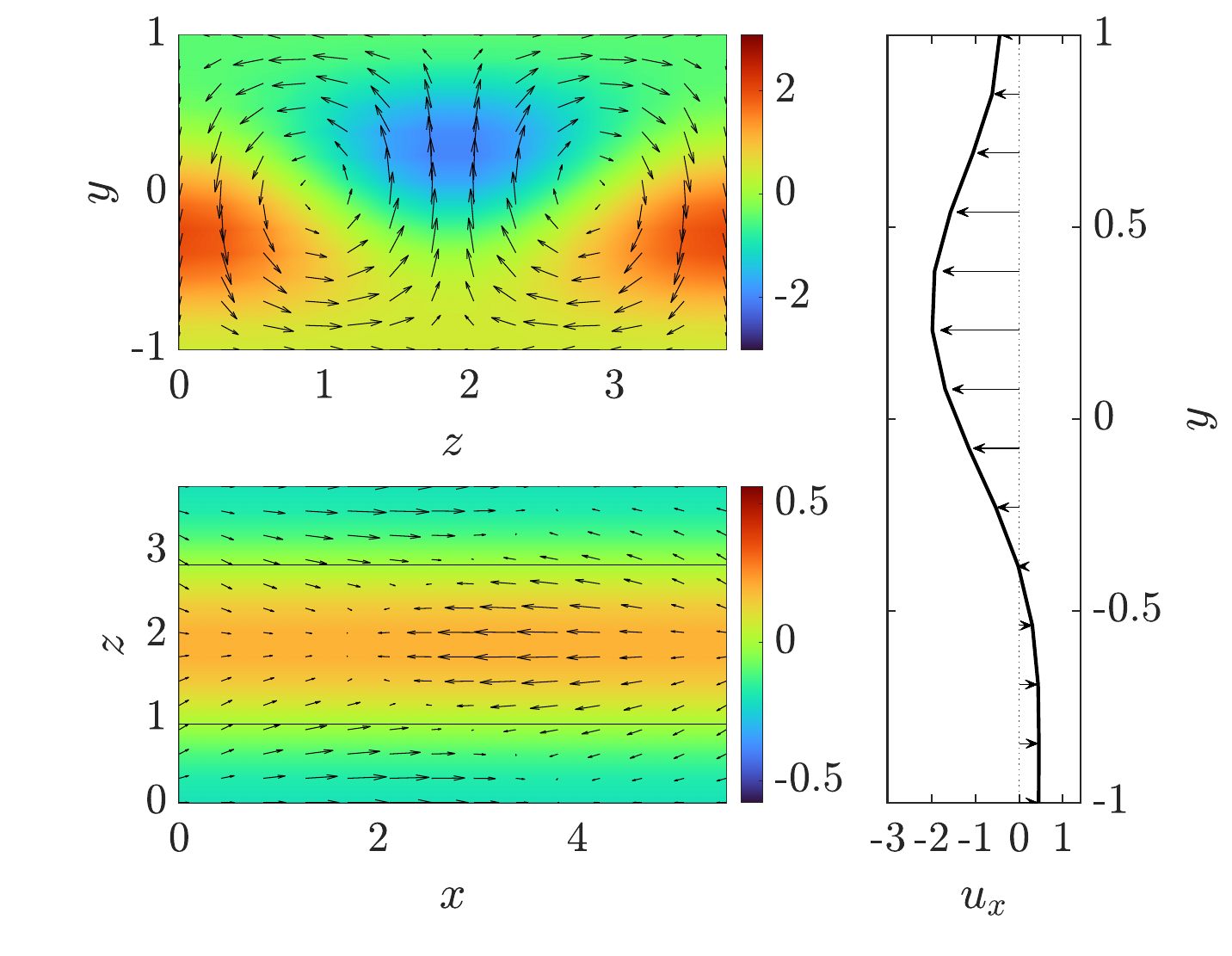}}
\subcaptionbox{$T=8$: Monotonic energy decay from this point $E(t)=E_0$.}{\includegraphics[width=0.45\textwidth]{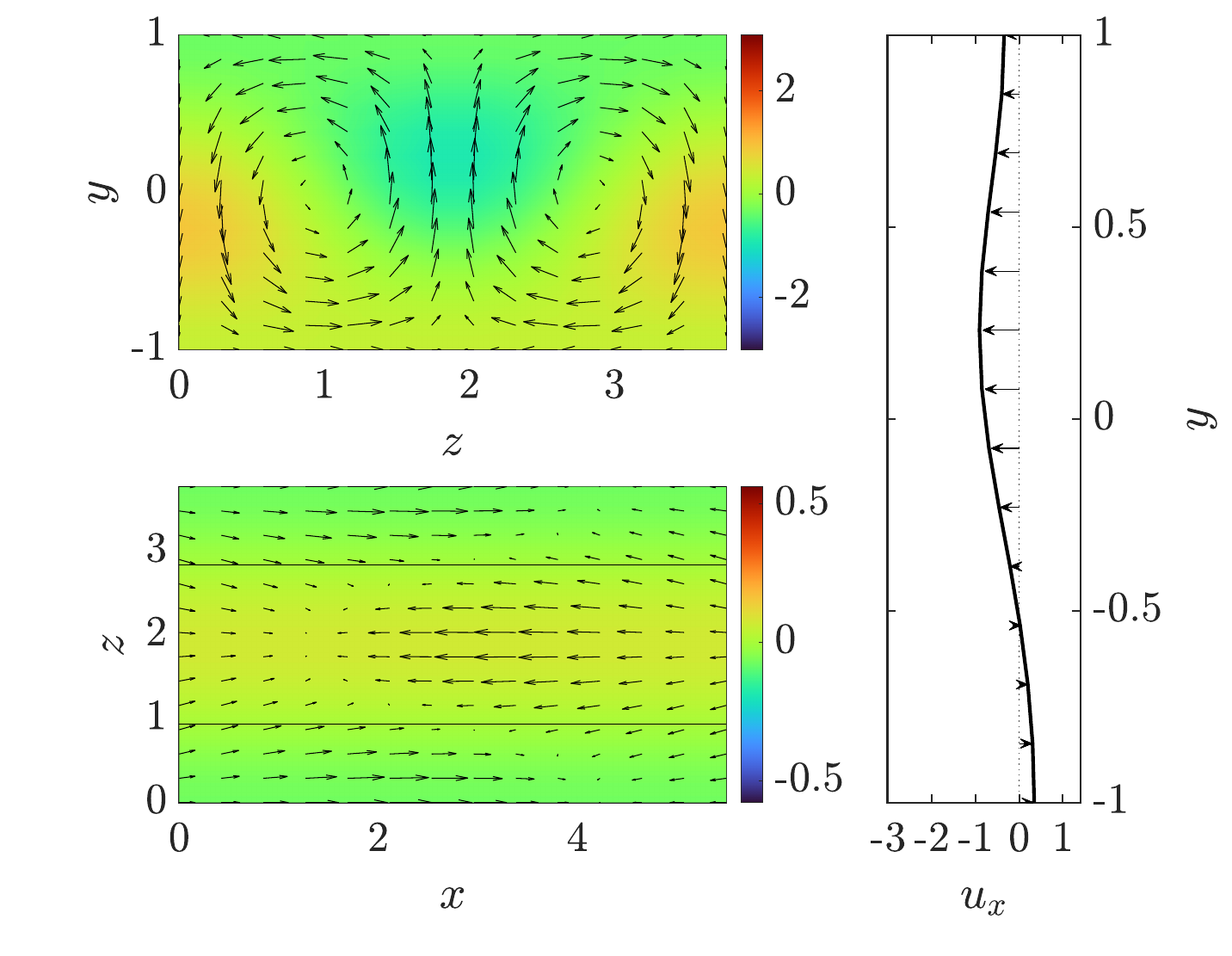}}
\caption{By maximizing the streak-formation at $t=0$, energy is maximized for time $T$, revealing a mechanism similar to the beginning of a turbulent `self sustaining' process.}
\label{fig:Flow_Timeline}
\end{figure*}

As shown in Figures \ref{fig:Flow_nonspVssp} and \ref{fig:Flow_Timeline}, maximizing mode $x_3$ in the optimal perturbation leads to the formation of strong stream-wise vortices. The stream-wise vortices in turn form streaks, which are clearly visible in the $x-z$ plots for $t>0$. This behavior is similar to the first step of the `self-sustaining process' \cite{hamilton_kim_waleffe_1995,Waleffe}, which consists of three phases: (1) formation of streaks by stream-wise vortices, (2) the breakdown of streaks, and (3) the regeneration of the stream-wise vortices. When a sufficiently high Reynolds number is considered, each stage preempts the next, resulting in a self-regenerating turbulent cycle. Although the process is not regenerative for the configuration examined here, features that resemble the transition are present. To show this, an overview of a $k=1$ flow with $d=d_0$ is given for several time steps in Figure \ref{fig:Flow_Timeline}. Here, the stream-wise velocity is amplified over the time interval $[0,T]$ which in the case of $k=1$ is solely the result of vortex and streak induced cross-stream disturbances. This behavior, known as the `lift-up' mechanism \cite{Ellingsen}, drives the flow to its maximum transient energy at time $T$. The lifting of the streak to form a locally high inflectional velocity profile at time $t=T$ resembles the disturbance growth mechanism that is key in the transition to turbulence \cite{Landahl_shearing,Brandt2014,Moehlis2004}.

Due to the non-convexity of the underlying optimization problem, the results reported in Figure \ref{fig:Sparsity} correspond to local optima.
During our investigation, multiple solution branches emerged for this system. By changing the seed from which sparse NLOP is performed, the optimal perturbations and associated $\sigma$ values change, as shown Figure \ref{fig:Nonuinque}. The existence of multiple solution branches was also encountered in~\cite{Foures2013}.
Interestingly, although the signs and exact values of the optimal perturbations in Figure \ref{fig:Nonuinque} differ from those in Figure \ref{fig:Sparsity}, the sparsity patterns between the two are similar. Furthermore, the maximum amplifications corresponding to each cardinality are the same as those presented in Figure \ref{fig:EvSP}. The flows generated using the optimal perturbations from Figure \ref{fig:Nonuinque} exhibit the same mechanisms discussed previously, though flipped about the coordinate axes. This flip makes sense as a property of the sinusoidal shear flow is its top-down symmetry. We note that the mean flow profile shape at $t=T$ is also the same which together with the previous results suggests that the flow is most sensitive around $y=\pm 0.2$ for $Re=20$.

For a given perturbation amplitude $E_0=d^2$, there is an optimal time horizon $T^*$ that results in the maximum transient growth over all time horizons (i.e.,~$T^*=\argmax_T\max_{\|x(0)\|_2^2=d^2} E(T)$).
From Figure \ref{fig:EvsD}, we observe that increasing the perturbation amplitude $E_0$ results in a smaller $T^*$, indicating a shorter time to the maximum transient growth.
This is a consequence of the higher $E_0$ resulting in stronger initial vortices, which in turn cause stronger streaks. These streaks are then lifted up over a shorter time window $T$, causing the velocity profile to reach its locally maximized inflection sooner for higher $E_0$.
Although we do not observe the onset self-sustaining process, these fundamental behaviors bear a strong resemblance to well-established mechanisms that lead to sustained turbulence in shear flows~\cite{Landahl_shearing,hamilton_kim_waleffe_1995,Moehlis2004}.

\section{Conclusions} \label{sec:conclusions}
In this work, we presented a framework for computing sparse finite-amplitude perturbations that maximize the transient growth of perturbation energy in nonlinear dynamic systems. 
The proposed approach can also be extended to compute spatially-localized perturbations in systems governed by partial differential equations.
A variational approach was used to formulate an iterative direct-adjoint looping algorithm to compute solutions to the underlying dynamically constrained optimization problem.
We first demonstrated the sparse NLOP framework on a simple 2-state model to develop an intuition for the approach and the underlying optimization problem.
The graphical analysis afforded by this simple example also facilitated the development of a simple heuristic for seeding the sparse NLOP problem using a linear estimate.
Subsequently, we applied the sparse NLOP framework to analyze a reduced-order (9-state) model of a sinusoidal shear flow at $Re=20$. 
Notably, the method identified the basic profile modifier mode---a core feature of the 9-state model~\cite{Moehlis2004}---as an important mode for driving instabilities.
In addition, the sparse NLOP framework
identified perturbations giving rise to the formation of streaks through strong downstream vortices and as the driver of instability.
Our observation is consistent with the literature and resembles the initial steps of the `self sustaining process' that drives recurring turbulence, which the model was constructed to predict~\cite{Moehlis2004}.
Our results suggest the framework outlined here will open new possibilities for the analysis and control of nonlinear flows.
In future work, we plan to implement sparse NLOP directly on the full Navier-Stokes equations to identify spatially-localized perturbations that give rise to pertinent nonlinear flow interactions that drive instabilities.
Sparse NLOP will also be employed to address questions pertaining to optimal actuator placement for active flow control applications.

\section*{Acknowledgments}
This material is based upon work supported by the Air Force Office of Scientific Research under award number FA9550-21-1-0106 and FA9550-21-1-0434,  and the National Science Foundation under grant number CBET-1943988.
An earlier version of this work was presented in AIAA Paper 2023-4258 \cite{AIAA_Paper}.

\bibliography{Bibliography}

\clearpage
\onecolumngrid

\begin{figure}[h!]
\subcaptionbox{$t=0$ with $k=1$ and $d=d_0$}{\includegraphics[width=0.45\textwidth]{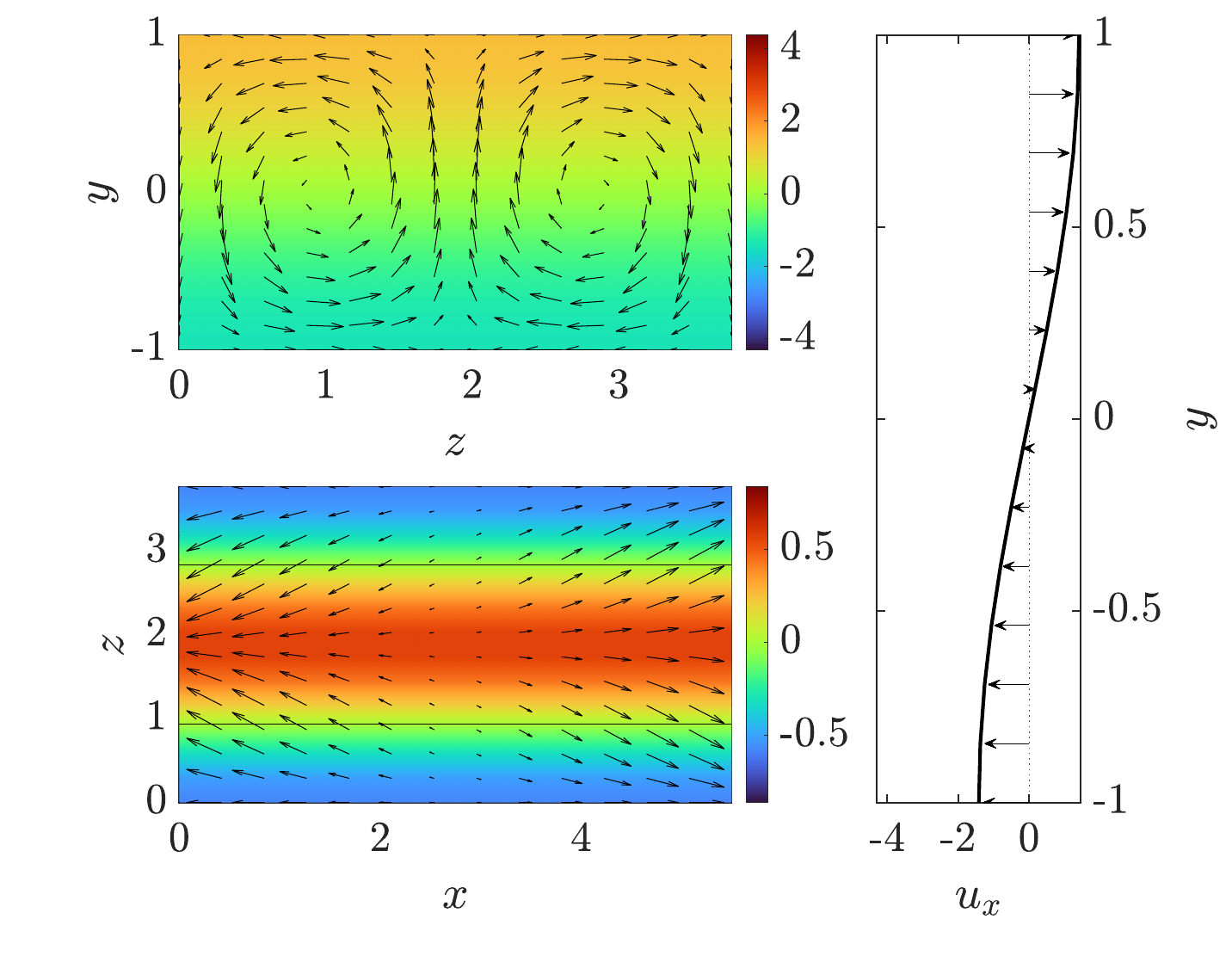}}
\subcaptionbox{$T=2.5$ with $k=1$ and $d=d_0$}{\includegraphics[width=0.45\textwidth]{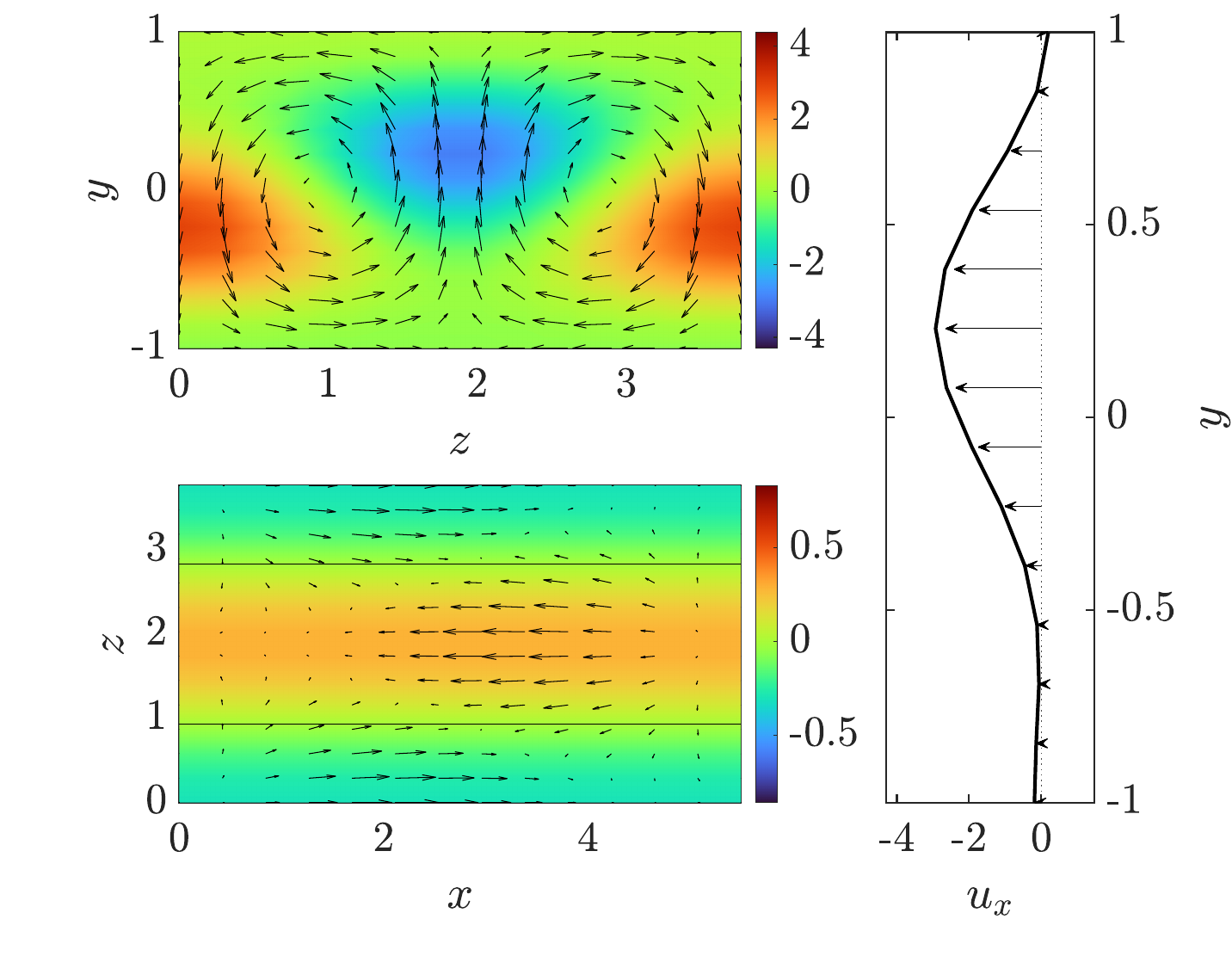}}
\caption{Velocity fields corresponding to the sparsest ($\sigma_1$) optimal perturbation from Figure \ref{fig:Sparsity} with perturbation size $d_0$. Within each block, the top left plots show the flow averaged in the streamwise direction, with the arrows indicating the $(y,z)$ plane flow and the color indicating the averaged x-velocity. The bottom left plots show the flow in the midplane between plates (at $y=0$). Here the arrows show the $(x,z)$ plane velocity and the colors indicate the $y$ velocity. Finally, the right plots shows the mean velocity profile.}
\label{fig:Flow_nonspVssp1}
\end{figure}

\begin{figure}[h!]
\subcaptionbox{$t=0$ with $k=4$ and $d=7d_0/4$}{\includegraphics[width=0.45\textwidth]{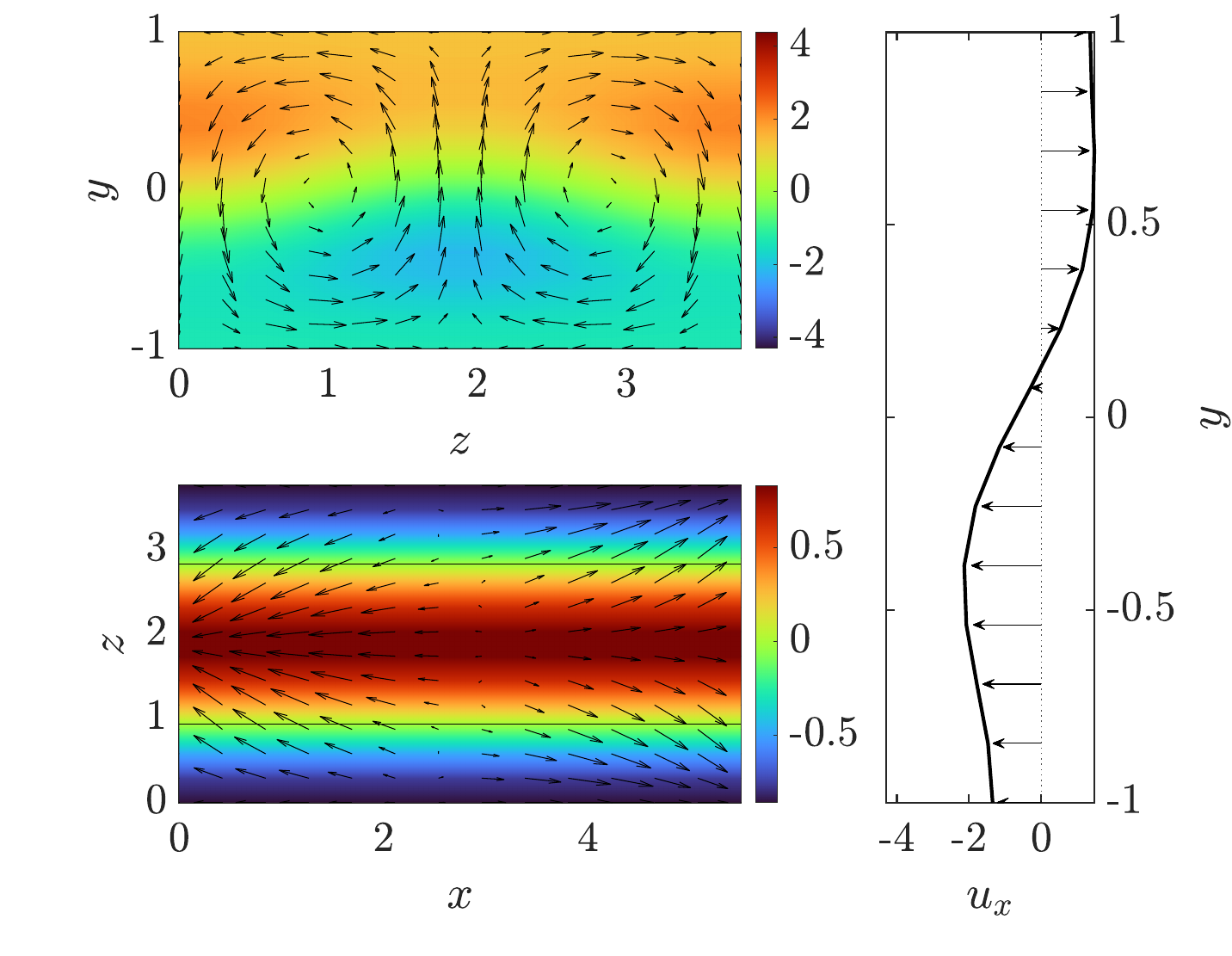}}
\subcaptionbox{$T=0.4$ with $k=4$ and $d=7d_0/4$}{\includegraphics[width=0.45\textwidth]{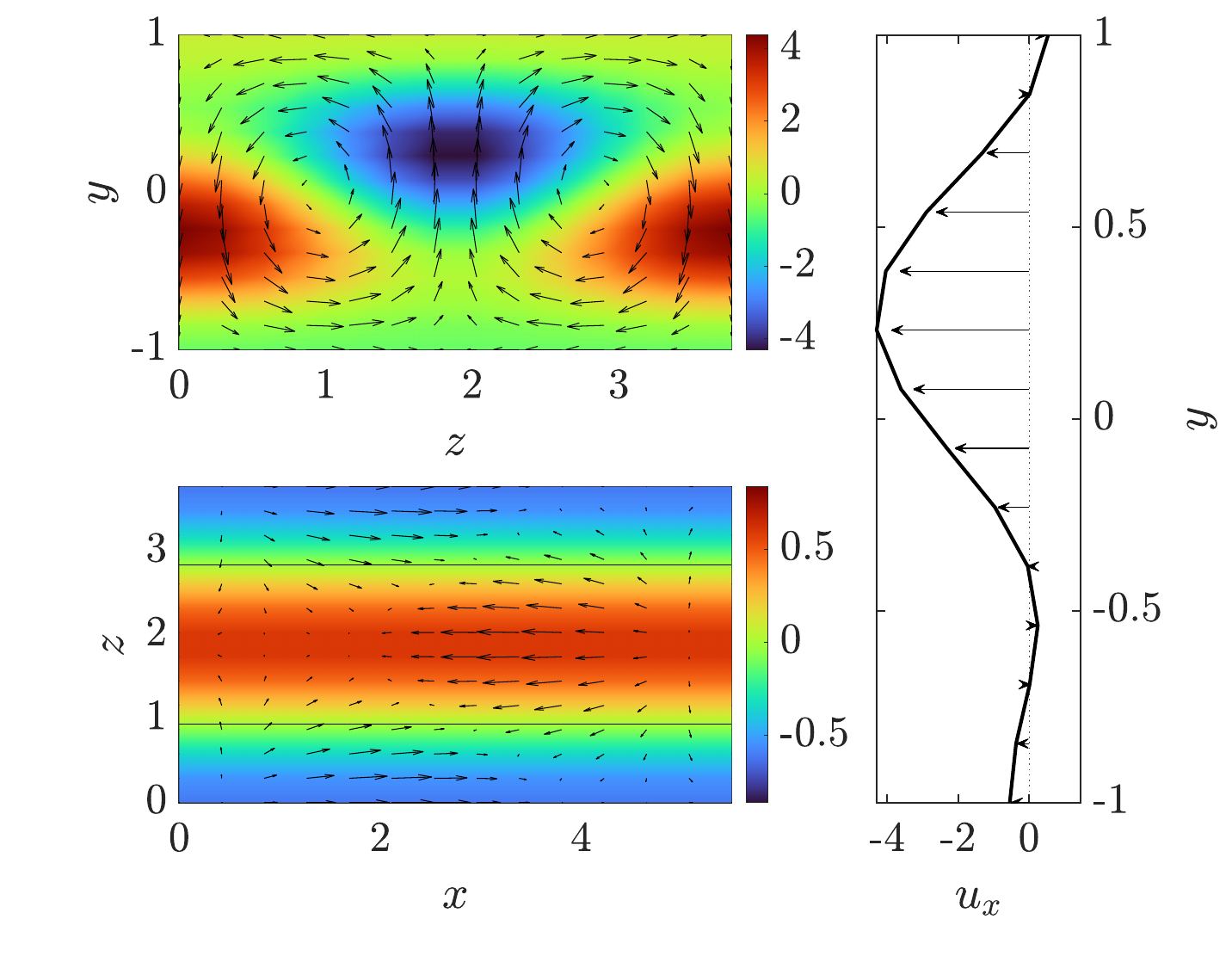}}
\caption{The flows at $T$ exhibit the same fundamental behavior for all $k$ and perturbation sizes that were shown in figure \ref{fig:EvsD}. Here, $k=4$ and the perturbation is $7d_0/4$. The lift-up mechanism is the same as seem in figure \ref{fig:Flow_nonspVssp1}, though due to the higher $E_0$ the magnitude of the disturbances are greater.}
\label{fig:Flow_nonspVssp}
\end{figure}

\begin{figure*}[h!]
\centering
\begin{subfigure}{0.8\textwidth}
    \includegraphics[width=\textwidth]{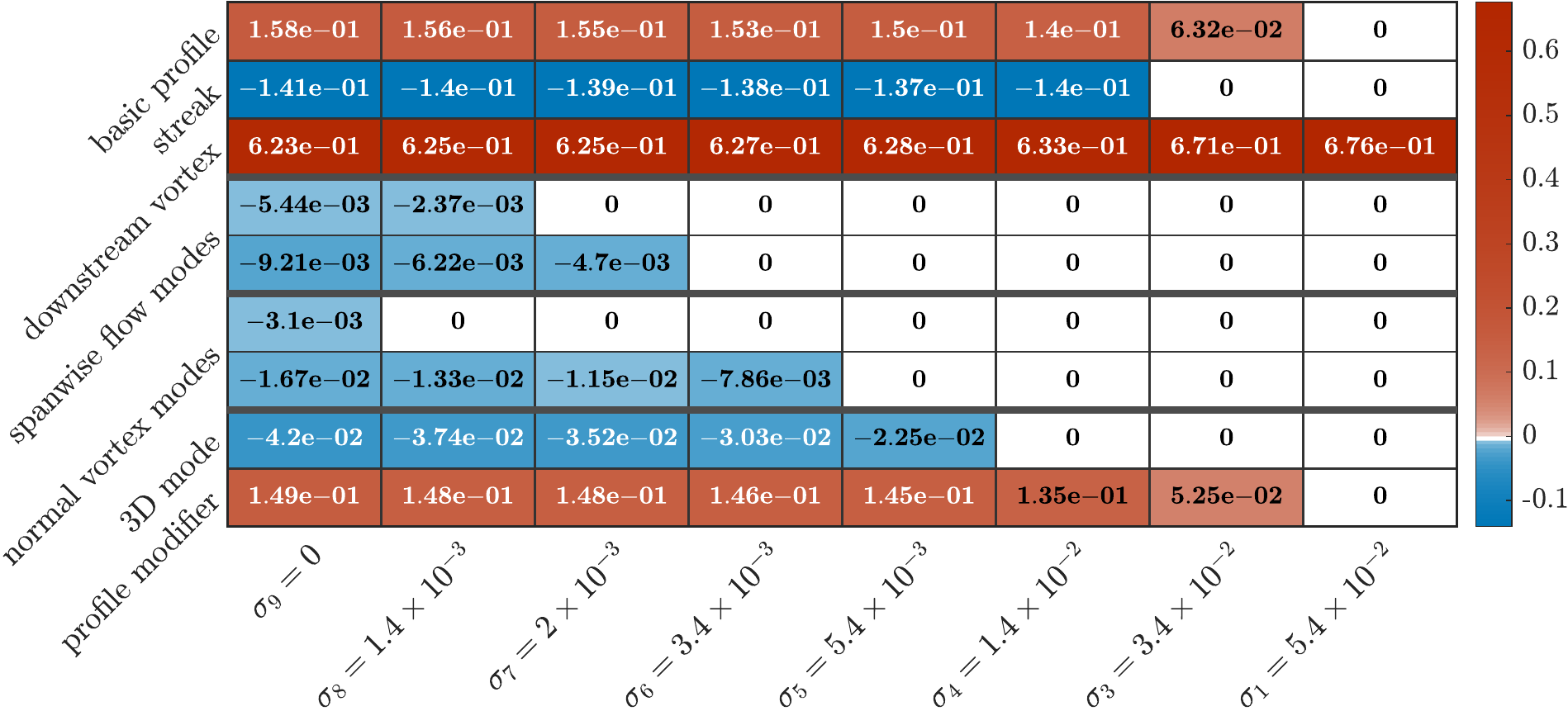}
    \caption{Just as for Figure \ref{fig:Sparsity},the optimal perturbations are found for the case where  $Re=20$, $T=2.5$, and $d_0\approx0.676$. However, the initial seed is different from that used to find Figure \ref{fig:Sparsity}.}
\end{subfigure}
\hfill
\begin{subfigure}{0.45\textwidth}
    \includegraphics[width=\textwidth]{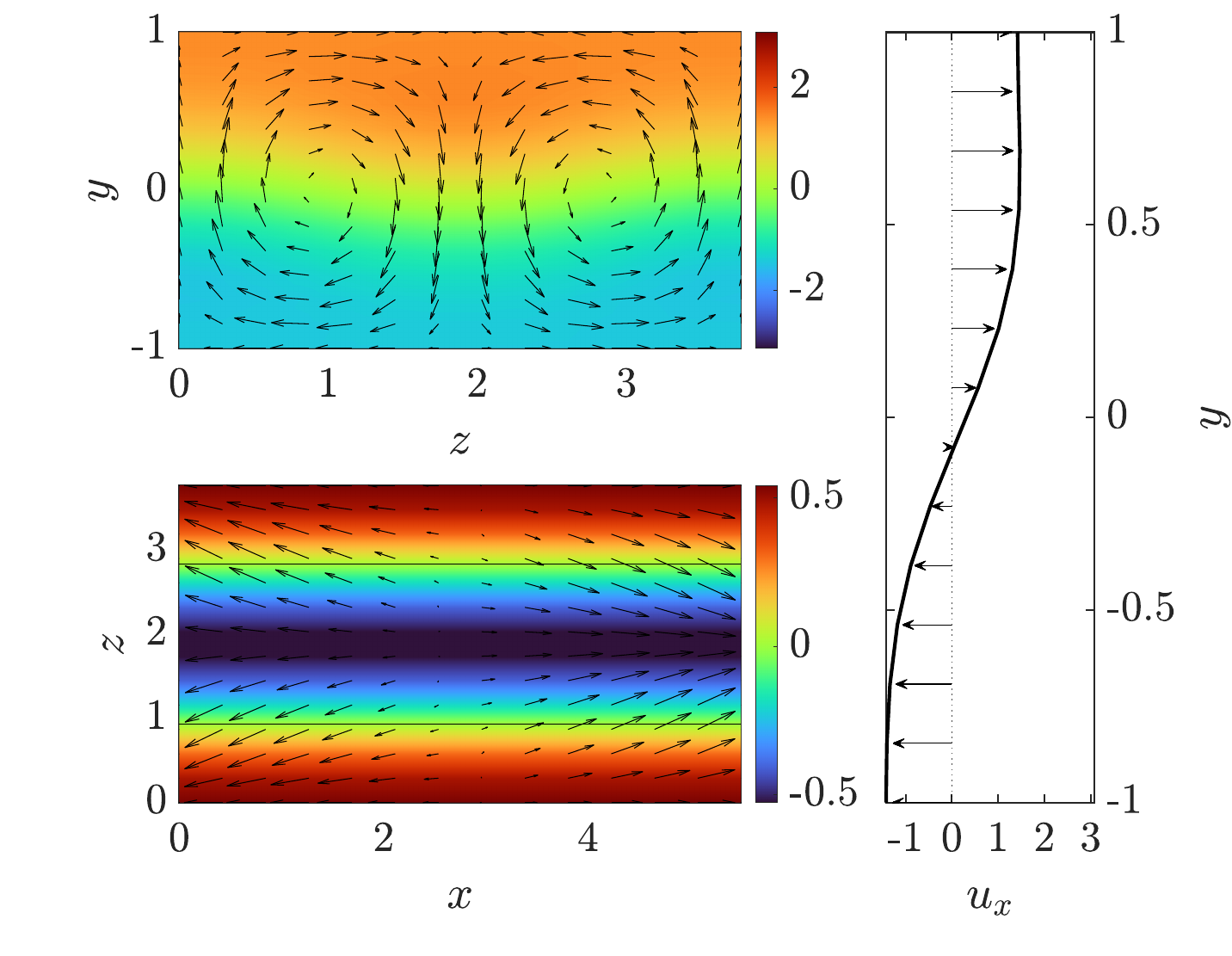}
    \caption{$t=0$ with $k=4$: Streak formation.}
\end{subfigure}
\begin{subfigure}{0.45\textwidth}
    \includegraphics[width=\textwidth]{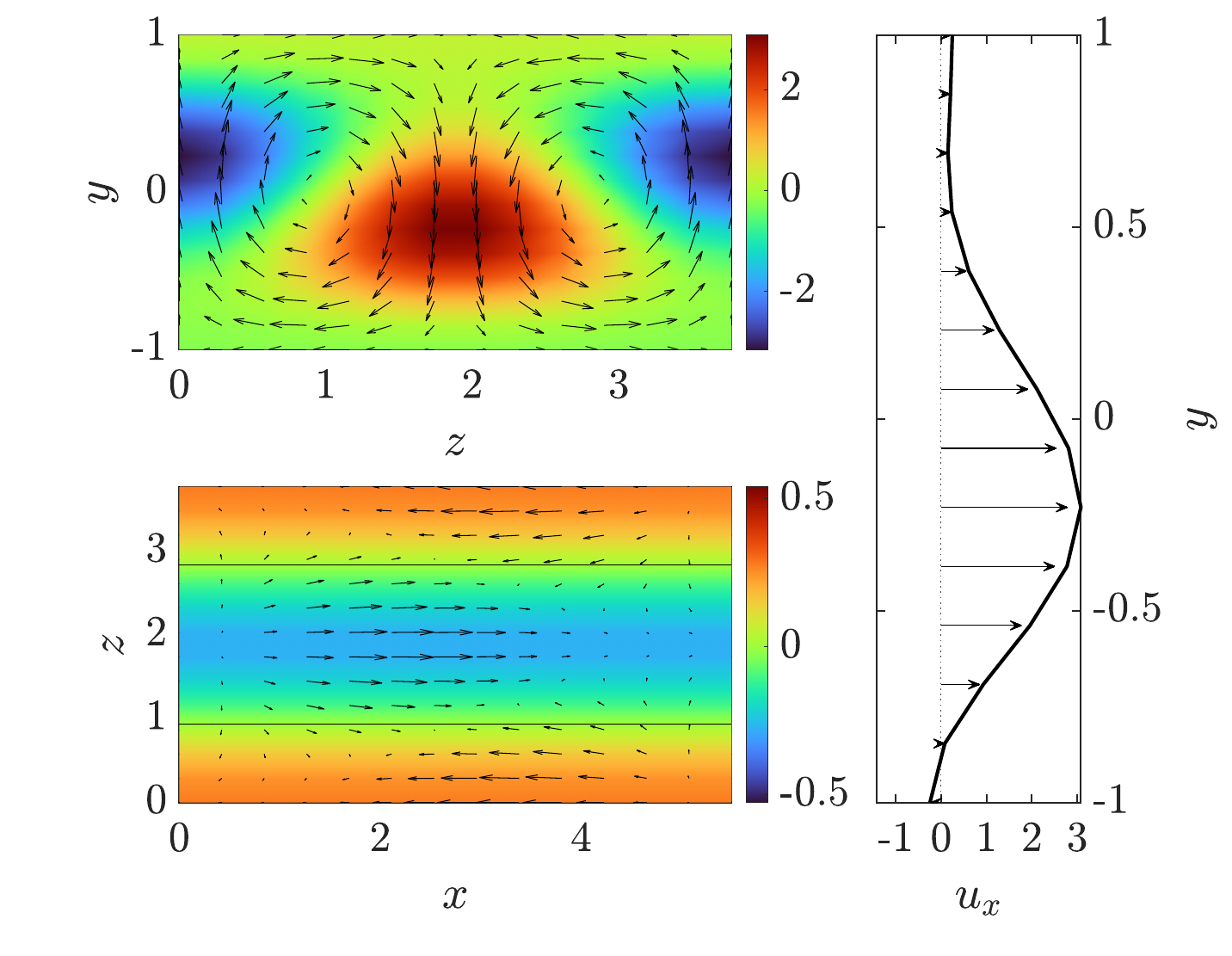}
    \caption{$T=2.5$ for $k=4$: $E(T)$ is maximized.}
\end{subfigure}
\caption{Even though the solutions to the sparse NLOP problem are non-unique, the energy-maximizing mechanism is the same. The streak formation and base profile deflection is the same as in Figures \ref{fig:Flow_Timeline} and \ref{fig:Flow_nonspVssp}, though it is flipped across $y=0$. The energies $E(T)$ associated with this result are the same as for the optimal perturbations in Figure \ref{fig:Sparsity}.}
\label{fig:Nonuinque}
\end{figure*}
\end{document}